\documentclass{cernrep}
\usepackage{amsmath,amssymb,amsfonts,color,graphicx,cite}
\graphicspath{{figs/}}

\newcommand{\ifb}{fb$^{-1}$}
\newcommand{\tb}{\tan\beta}

\newcommand{\gev}{\,\, \mathrm{GeV}}

\newcommand{\mgl}{m_{\tilde g}}
\newcommand{\msq}{m_{\tilde q}}
\newcommand{\cha}[1]{\tilde \chi^\pm_{#1}}

\newcommand{\neu}[1]{\tilde \chi^0_{#1}}

\title{Benchmark Models, Planes, Lines and Points for Future SUSY Searches at the LHC}
\author{
S.S.~AbdusSalam$^1$, 
B.C.~Allanach$^2$,
H.K.~Dreiner$^3$, 
J.~Ellis$^{4,5}$,
U.~Ellwanger$^{6}$,
J.~Gunion$^{7}$,
S.~Heinemeyer$^8$,
M.~Kr\"amer$^9$, 
M.~Mangano$^5$,
K.A.~Olive$^{10}$,
S.~Rogerson$^{11}$,
L.~Roszkowski$^{12,13}$, 
M.~Schlaffer$^{14,15}$,
G.~Weiglein$^{15}$
}
\institute{
$^1$ The Abdus Salam International Centre for Theoretical Physics, Strada Costiera 11, I-34014 Trieste, Italy\\
$^2$ Department of Applied Mathematics and Theoretical Physics,
     Wilberforce Road, Cambridge University, Cambridge
     CB3 0WA, UK\\
$^3$ Bethe Center for Theoretical Physics \& Physikalisches Institut der
     Universit\"at Bonn, Germany\\
$^4$ Theoretical Particle Physics and Cosmology Group, Department of Physics, 
     King's College London, London WC2R 2LS, UK\\
$^5$ CERN, CH--1211 Gen\`eve 23, Switzerland\\
$^{6}$ Laboratoire de Physique Th\'eorique, UMR 8627, CNRS and
  Universit\'e de Paris--Sud, B\^at. 210, F-91405 Orsay, France \\
$^{7}$ Department of Physics, University of California, Davis, CA
95616, USA \\
$^8$ Instituto de F\'{\i}sica de Cantabria (CSIC-UC), E--39005
     Santander, Spain\\
$^9$ Institute for Theoretical Particle Physics and Cosmology, RWTH Aachen
     University, D-52056 Aachen, Germany\\ 
$^{10}$ William I. Fine Theoretical Physics Institute \& Department of Physics,
University of Minnesota, Minneapolis, MN 55455, USA\\
$^{11}$ High Energy Physics Group, Blackett Laboratory, Imperial College,
Prince Consort Road, London SW7 2AZ, UK\\ 
$^{12}$ National Centre for Nuclear Research, Hoza 69, 00-681 Warsaw, Poland\\
$^{13}$ Department of Physics and Astronomy, University of Sheffield, 
     Sheffield S3 7RH, UK\\
$^{14}$ Ludwig--Maximilians--Universit\"at M\"unchen, Fakult\"at f\"ur Physik,
Arnold Sommerfeld Center for Theoretical Physics, D-80333 M\"unchen,
Germany\\
$^{15}$ DESY, Notkestra\ss{}e 85, D--22607 Hamburg, Germany
}

\begin{document}
\maketitle 
\begin{abstract}
We define benchmark models for SUSY searches at the LHC, including the 
CMSSM, NUHM, mGMSB, mAMSB, MM-AMSB and p19MSSM, as
well as models with R-parity violation and the NMSSM.
Within the parameter spaces of these models, we propose benchmark subspaces, 
including planes, lines and points along them. The planes may be useful
for presenting results of the experimental searches in different SUSY
scenarios, while the specific benchmark points may serve for more
detailed detector performance tests and comparisons. We also describe
algorithms for defining suitable benchmark points along the proposed
lines in the parameter spaces, and we define a few benchmark points 
motivated by recent fits to existing experimental data.
\end{abstract}

\noindent
\begin{center}
{\small CERN-PH-TH/2011-224, DAMTP-2011-69, DESY 11-90, FTPI-MINN-11/22, \\
KCL-PH-TH/2011-30, LCTS/2011-15, LPT Orsay 11-76, TTK-11-41, UMN-TH-3012/11}
\end{center}

\section{Introduction}
\label{sec:intro}

Softly broken low-energy supersymmetry (SUSY) has many attractive
features \cite{mssm}. For example, unlike the Standard Model (SM), it
provides an elegant mechanism for stabilizing the gauge hierarchy with
respect to the effects of radiative corrections and a natural
weakly-interacting dark matter (DM) candidate, in addition to
facilitating gauge coupling unification, predicting the existence of a
light Higgs boson and potentially making a desirable contribution to
the anomalous magnetic moment of the muon, $(g - 2)_\mu$. On the other
hand, SUSY has to be (softly) broken in order to make contact with
reality, which even in 
the general 
Minimal Supersymmetric Standard Model (MSSM) 
introduces a large number of new free
parameters, such as the soft SUSY-breaking (SSB) masses, mixings and
complex phases in the couplings. Because of
SUSY's natural link with grand unified theories (GUTs), one often
explores SUSY models in which various boundary conditions are imposed
on the SSB parameters at the GUT scale.

ATLAS and CMS have already made great strides in searches for
SUSY with $\sim 1$~\ifb\  of data each~\cite{Aad:2011kz,Aad:2011xm,Aad:2011xk,Aad:2011ks,Aad:2011yf,daCosta:2011qk,Aad:2011hh,Chatrchyan:2011-new,Chatrchyan:2011qs,Chatrchyan:2011ek,Collaboration:2011ida,Chatrchyan:2011bj,Chatrchyan:2011ff,Chatrchyan:2011ah,Chatrchyan:2011wb,Chatrchyan:2011bz,Chatrchyan:2011wc,Khachatryan:2011tk,Khachatryan:2010uf}.
However, the presentation of the experimental results from searches for
supersymmetric particles or other kinds of new physics at the LHC
necessarily involves a certain dependence on the assumed model of new
physics. Results from SUSY searches are usually presented either within
the parameter space of a specific SUSY model or as cross section limits
or limits on the masses of particles of ``simplified
models''~\cite{Alves:2011wf}.  While the
latter approach provides in principle information that is less 
model-dependent, the former approach has the advantage that the specific 
SUSY models can be confronted also with other constraints, such as limits
from Higgs searches and searches for dark matter as well as
constraints from electroweak precision data and flavour physics.

In order to allow the interpretation of the experimental results in
terms of different possible manifestations of SUSY it is useful to
consider certain benchmark models that have fewer free parameters
than the most general incarnation of the MSSM. For this purpose we
consider several models with GUT-scale boundary
conditions on the SSB parameters that have frequently been studied up
to now in the context of collider
and dark matter searches. As an alternative, we
also consider a 
version of the MSSM with 19
parameters, as well as models with R-parity violation and the NMSSM. 
Within those models we propose some benchmark subspaces
that could be useful for the presentation of experimental results. 
We also define specific benchmark points that illustrate different possible
experimental signatures and may serve for more
detailed detector performance tests and comparisons. We choose those 
benchmark points along certain lines in the parameter spaces that share
distinctive experimental signatures, which may differ from those in the models
most commonly studied to date. If the
SUSY exclusion limits advance, one can move to the next benchmark point
along any of these lines. This could be useful for investigating the prospective
experimental sensitivity or for optimising the search in the relevant
parameter region, etc.


\section{Definition of models}
\label{sec:models}

\subsection{The Constrained MSSM (CMSSM)}
\label{sec:cmssm}

A commonly studied model within this broad class of GUT-based models is the Constrained MSSM
(CMSSM)~\cite{cmssm}, in which not only gaugino soft masses unify
to a common value $m_{1/2}$ at the GUT scale, but also the
SSB masses of all the sfermions and Higgs
doublets unify to a common value $m_0$. These parameters, along with a common
tri-linear SSB parameter $A_0$,
the ratio of Higgs vacuum expectation values $\tb$ and the sign of the
Higgs mixing parameter, sign($\mu$), 
form the four continuous and one discrete parameters of the model:
\begin{align}
\mbox{CMSSM:}\quad m_{1/2}, m_0, A_0, \tb, \rm{sign}(\mu)~.
\label{cmssm}
\end{align}
The unified parameters $m_{1/2}$, $m_0$, $A_0$ are thought to appear
via some gravity-mediated mechanism, and are defined at the GUT scale
$M_{\rm GUT} \approx 2 \times 10^{16} \gev$, whereas $\tb$ (and
sign$(\mu)$) are defined at the electroweak (EW) scale.  The CMSSM is
assumed to conserve multiplicatively the discrete R-parity,
  $R_P\equiv(-\mathbf{1})^{2S+L+3B}$, where $S$ is the spin of the
  particle, $L$ the lepton number and $B$ the baryon number.  This
ensures that the lightest supersymmetric particle (LSP) is stable and
a natural cold dark matter (CDM) candidate, often thought to be the lightest
neutralino.  The CMSSM, like many other simple unified models, is a
Minimal Flavor-Violating (MFV) scenario with no additional
flavor-violating terms beyond those in the SM.


\subsection{The Minimal Supergravity Model (mSUGRA)}
\label{sec:msugra}

Additional assumptions may be imposed beyond those in the CMSSM. For
example, in the minimal supergravity model (mSUGRA) there is a
specific relation between the trilinear and bilinear SSB parameters
and the universal scalar mass: $A_0 = B_0 + m_0$~%
\footnote{It should be noted that many publicly available codes use a
  different sign convention, $A_0 = B_0 - m_0$, as may be ascertained
  by comparing the signs of the gauge and Yukawa contributions to the
  renormalization group equations of the $A$ parameters. More details can be found
  in~\cite{mc4}.}%
~and the gravitino mass $m_{3/2}$ is fixed to be equal to the common scalar mass
before renormalization, $m_{3/2} = m_0$~\cite{msugra}.  Hence, the
model has just three free continuous parameters,
\begin{align}
\mbox{mSUGRA:} \quad m_{1/2}, m_0, A_0, \rm{sign}(\mu)~,
\label{vcmssm}
\end{align}
and $\tb$ is now fixed by the radiative
electroweak symmetry breaking conditions.

Relaxing the condition on the gravitino mass leads to the 
Very Constrained MSSM (VCMSSM)~\cite{vcmssm}, which has the same 
set of free parameters as mSUGRA but may evade the
restrictive cosmological and astrophysical
constraints due to late decays of neutralinos into gravitinos~\footnote{The conditions (\ref{vcmssm}) are often
applied at the unification scale $M_{\rm GUT}$, though the underlying theory may
enforce them at some other scale such as the Planck scale, and renormalization
of the SSB parameters between this scale and $M_{\rm GUT}$ could be significant.}.


\subsection{Non-Universal Higgs Mass Model (NUHM)}
\label{sec:nuhm}

Because of its economy, the CMSSM (and {\it a fortiori} more restricted versions) may be
missing some features of unified models with less restrictive boundary
conditions at the unification scale.  In particular, the assumption of
unification of the SSB Higgs mass with those of the sfermions may easily be relaxed
without prejudice to MFV
~\cite{nuhm1,nuhm2,nuhm_global}. 
Depending on whether the two soft SSB parameters of the Higgs
sector are the same (NUHM1) or not (NUHM2, often simply termed the NUHM) the free
parameters of this model are
\begin{equation}
\label{nuhm1}
\mbox{NUHM1:}\quad m_{1/2}, m_0, m_H, A_0, \tb, \rm{sign}(\mu)~,
\end{equation}
where $m_H$ denotes the unified SSB parameter in the
Higgs sector at $M_{\rm GUT}$, or
\begin{equation}
\label{nuhm2}
\mbox{NUHM2:}\quad m_{1/2}, m_0, m_{H_u}, m_{H_d}, A_0, \tb, \rm{sign}(\mu)~,
\end{equation}
where $m_{H_u}$ and $m_{H_d}$ denote the two independent soft
SSB parameters in the Higgs sector at $M_{\rm GUT}$.

Equivalently, one can trade the new GUT scale parameters for one or two
parameters at the EW scale. In the NUHM1, $m_H$ can be traded for $M_A$
{\em or} $\mu$, whereas in the NUHM2 both $m_{H_u}$ and $m_{H_d}$ can be
traded for $M_A$ {\em and} $\mu$ as free parameters.


\subsection{Minimal Gauge-mediated SUSY breaking (mGMSB)}
\label{sec:gmsb}

The gauge-mediated SUSY-breaking (GMSB) model is constructed from a
GUT-scale scenario for SUSY-breaking mediation in which
communication to the visible sector is via gauge
 interactions~\cite{IR,Giudice:1998bp}. 
Minimal GMSB (mGMSB) models have four
continuous parameters, namely the messenger field mass-scale,
$M_{\rm mess}$, the visible-sector SSB scale, $\Lambda$,
$\tan \beta$, $c_{\rm grav} \geq 1$ (a factor for the gravitino mass) 
as well as a discrete parameter, $N_{\rm mess}$, representing
the number of $SU(5)$ representations of the mediating fields:
\begin{align}
\mbox{mGMSB:}\quad M_{\rm mess}, \; \Lambda, \; \tb, \; c_{\rm grav},\; N_{\rm mess}. 
\end{align}
The next-to-lightest supersymmetric particle (NLSP) plays an important role in
the 
phenomenology, since cascade decay chains of each produced sparticle will
typically end in the NLSP.  
It is commonly the stau, which decays to a tau plus gravitino,
or a neutralino, which often decays to a photon plus a gravitino. 
Depending on the parameters, the NLSPs can range between being stable on the
time-scales taken to cross a detector, to being prompt and decaying at the
interaction point. 
For example, in the case of a neutralino NLSP, 
its decay length is approximately given by~\cite{Ambrosanio:1997rv}
\begin{equation}
L_{\mbox{decay}} =  \frac{1}{\kappa_{\gamma}} \left(
\frac{100~\mbox{GeV}}{m_{\rm NLSP}} \right)^5 
\left(  \frac{\Lambda}{100~\mbox{TeV}}  \right)^2 
\left(  \frac{M_{\rm mess}}{100~\mbox{TeV}}  \right)^2   10^{-4}~\mbox{m}, 
\label{length}
\end{equation}
where $\kappa_{\gamma}$ is the photino component of the neutralino. Here $\Lambda$
sets the sparticle mass scale, and varying
$M_{\rm mess}$ changes the decay length according to
Eq.~\ref{length}. We include in our list of proposed benchmarks
below a model with a quasi-stable neutralino
NLSP. According to 
Eq.~\ref{length}, this can be arranged by increasing $M_{\rm mess}$.
If $M_{\rm mess}$ is close to the GUT scale, the resulting spectra can
closely resemble 
those of the CMSSM~\cite{Allanach:2011ya}, but an intermediate $M_{\rm mess}$
can yield spectra that significantly differ from typical CMSSM ones. 
Some similar
features may be found in models that assume universal SSB parameters at
a scale below the GUT scale~\cite{subGUT}.


\subsection{Anomaly-mediated SUSY breaking (AMSB)}
\label{sec:amsb}

The anomaly-mediated SUSY-breaking (AMSB) model is constructed from a
GUT-scale scenario for SUSY-breaking mediation in which
communication to the visible sector arises from the super-Weyl
anomaly~\cite{Randall:1998uk,Giudice:1998xp}. In 
minimal AMSB (mAMSB) models there are three parameters: a parameter, $m_0$,
contributing to the squared scalar masses at the GUT scale which would
otherwise be negative; the vacuum expectation value of the supergravity field
representing the sparticle mass scale, $m_{\rm aux}$, and $\tan \beta:$
\begin{align}
\mbox{mAMSB:}\quad m_0, \; m_{\rm aux}, \; \tb~.
\end{align}


\subsection{Mixed modulus-anomaly-mediated SUSY breaking (MM-AMSB)}
\label{sec:mm-amsb}
 
Mixed modulus-anomaly-mediated SUSY breaking scenarios are
inspired by models of string compactification with
fluxes~\cite{Kachru:2003aw}. In these scenarios the soft terms receive
contributions from both gravity-mediated and anomaly-mediated SUSY
breaking~\cite{Choi:2005ge}, whose relative sizes are characterized by a phenomenological parameter
$\alpha$. This mixed scenario is also known as ``mirage mediation",
since it appears that the gaugino masses
unify at an intermediate scale
well below $M_{\rm GUT}$. 
MM-AMSB models are specified by the parameters
\begin{align}
\mbox{MM-AMSB:}\quad m_{3/2}, \alpha, \tb, {\rm sign}(\mu), n_i, l_a~,
\label{mmamsb}
\end{align}
where 
$\alpha$ determines the relative weight of anomaly and gravity
mediation, the $n_i$ are the modular weights of the visible sector
matter fields, and $l_a$ appears in the gauge kinetic
function~\cite{Baer:2006id}.  In MM-AMSB models the relative size of
the gaugino masses $M_1, M_2$ and $M_3$ is determined by the parameter
$\alpha$, and different values of $\alpha$ thus correspond to
different mass patterns for the strongly- and weakly-interacting
gauginos.


\subsection{The phenomenological 19-parameter MSSM (p19MSSM)}
\label{sec:pmssm}
A complementary framework to the above GUT-scale models for SUSY
breaking is the phenomenological MSSM, with all of its many free
parameters specified at the electroweak scale. In order to make a
phenomenological analysis of its vast parameter space manageable,
various well-motivated simplifying assumptions have been made.  On the
other hand, some of the main assumptions used in the previous sections
in reducing the number of MSSM parameters to construct the GUT-scale
models, namely, the physics behind SUSY breaking, the mediation
mechanism and the renormalization group (RG) running of the parameters
from the SSB scale are not relevant for, and hence decoupled from,
typical MSSM parametrizations.

One MSSM parametrization that has been recently extensively studied
involves 19 free parameters of the MSSM
(p19MSSM)~\cite{Djouadi:2002ze,AbdusSalam:2008uv,Berger:2008cq,p19mssm}.  
In this approach, the resulting number of SSB parameters 
are derived from the
parent 105 parameters of the MSSM by removing all sources of
CP violation and generation mixing beyond the SM: assuming the
parameters to be real, that all off-diagonal elements in the
sfermion mass matrices 
are equal to zero, 
 and that first- and second-generation soft terms are equal. Only the
 trilinear couplings most relevant for SUSY effects,  
$A_t$, $A_b$, $A_\tau$ 
are included as free parameters. The other $A_{f\neq t,b,\tau}$ are set to zero.

The free parameters of the p19MSSM model are 
\begin{align}
\mbox{p19MSSM:} &\quad  
M_1, M_2, M_3; \; \nonumber \\
& m_{\tilde u_L}=m_{\tilde d_L}=m_{\tilde c_L}=m_{\tilde s_L}, \;
m_{\tilde u_R}=m_{\tilde c_R}, \; m_{\tilde d_R}=m_{\tilde s_R}, \;
m_{\tilde e_L}=m_{\tilde \mu_L}, \; m_{\tilde e_R}=m_{\tilde \mu_R};\; 
\nonumber \\
& m_{\tilde t_L}=m_{\tilde b_L}, \; m_{\tilde t_R}, \; m_{\tilde b_R}, \;  
m_{\tilde \tau_L}, \; m_{\tilde \tau_R}; \nonumber \\
& A_t, A_b, A_\tau; \quad \mu, M_A, \tb
\end{align}
where $M_{1,2,3}$ are the SSB parameters in the
gaugino and gluino sector,
$m_{\tilde f_L}$ and $m_{\tilde f_R}$ are the diagonal SSB parameters
in the sfermion sector ($f=u,d,c,s,t,b,e,\mu,$ or $\tau$), and
$A_{f=t,b,\tau}$ denote the
trilinear Higgs-sfermion coupling in the third generation.
An alternative choice of parameterisation would be to replace 
the CP-odd neutral Higgs mass parameter, $M_A$, and the 
Higgs doublet mixing parameter, $|\mu|$, by the two SSB
parameters in the Higgs sector, $m^2_{H_u}$ and $m^2_{H_d}$. 

Other versions of the MSSM with different choices of simplifying
assumptions, leading to either a reduced or enlarged number of free
parameters, have also been extensively studied in the literature. For
some recent work done in the context of the LHC, see,
\textit{e.g.}~\cite{Allanach:2011ya,mssm-recent,Conley:2011nn,AbdusSalam:2010qp,AbdusSalam:2011hd}.


\subsection{MSSM with RPV}
 \subsubsection{Definition of the RPV-CMSSM}

As stated above, in the CMSSM R-parity $R_p$ is assumed to hold.
Initially, the justification for this symmetry was that it
guarantees the stability of the proton, and another attractive feature is that it leads to 
a good dark matter candidate such as the lightest
neutralino.

In the RPV--MSSM the discrete symmetry R-parity is replaced by a
different symmetry (such as baryon--triality, or lepton-parity), which
guarantees proton stability but allows for either lepton--number
violation or baryon--number violation. The LSP is then not stable, and
can decay inside or outside the detector, depending on the size of the
coupling. Another particle such as the axion, axino or gravitino~\cite{OLR} must then
comprise the dark matter in the universe. On the other side, the
advantage is that the neutrino masses are automatically included and
naturally light \cite{Allanach:2003eb}.

The extra possible couplings in the superpotential are
\begin{equation}
W_{\rm RPV}=\epsilon_{ab}\left[\frac{1}{2}\lambda_{ijk}  L_i^aL_j^b E^C_k+
\lambda'_{ijk} L_i^aQ_j^b D^C_k+\kappa_i L_i^a H_2^b\right]
+\frac{1}{2}\epsilon_{xyz}\lambda_{ijk}'' U_i^{xC} D_j^{yC} D_k^{zC}\,, 
\label{eq:rpv}
\end{equation}
where $i,j,k\in\{1,2,3\}$ are generation indices, $a,b\in\{1,2\}$
and $x,y,z\in\{1,2,3\}$ are SU(2) and SU(3) fundamental representation
indices, respectively, and $^{C}$ is the usual charge-conjugation. 
The parameters of the constrained RPV--MSSM
(RPV-CMSSM; corresponding to the CMSSM) are \cite{Allanach:2003eb}
\begin{align}
\mbox{RPV--CMSSM:}\quad m_{1/2},\, m_0,\, A_0,\, \tb,\,\rm{sign}(\mu),
\,\Lambda~,
\label{rpv--cmssm}
\end{align}
where the CMSSM parameters are all defined at the unification scale, and
$\Lambda$ corresponds to exactly one of the R-parity violating
parameters
\begin{align}
\Lambda\in\{\lambda_{ijk},\,\lambda'_{ijk},\,\lambda''_{ijk},\,\kappa
_i\}\,.
\end{align}
Note that in the CMSSM version discussed in~\cite{Allanach:2003eb},
$\Lambda$ was defined at $M_{\rm GUT}$,
whereas here we
define it at $M_Z$, in order to make the analysis of the
possible topologies easier.
For the bilinear terms $\kappa_i L_iH_u$,
$\kappa_i(M_{\mathrm{GUT}})=0$ at the unification scale. There are
however other models that have just one bi-linear term, non--zero at
the unification scale: $\Lambda=\kappa_i\not=0$: see for example
Ref.~\cite{Hirsch:2008dy}. 
Note furthermore, that even though just one
RPV--coupling is non--zero at the unification scale in these models,
the RGEs generate other non--zero couplings at the weak scale. 

Due to the diverse phenomenology that
may arise from the RPV couplings, we summarize here its
general features~\footnote{The RPV-CMSSM as described is programmed in \texttt{SOFTSUSY}, with the  
SUSY Les Houches accord option for inputting $\Lambda(M_Z)$ \texttt{BLOCK 
  SOFTSUSY} parameter \texttt{8}, as described in the
\texttt{SOFTSUSY} 
manual~\cite{Allanach:2009bv}. }.

\subsubsection{Phenomenology of the RPV MSSM}
In the RPV MSSM the LSP typically decays promptly in the detector for
couplings larger than about $10^{-6}$.  The resulting number of
different possible new signals coming from the RPV MSSM is large,
since there are 48 new couplings, see Eq.~(\ref{eq:rpv}).
Furthermore, one can obtain single sparticle production, unlike in the
R-parity conserving case, if one assumes that $\lambda'_{ijk}$ or
$\lambda''_{ijk}$ is at least larger than about $3\cdot10^ {-3}$, for
first-generation incoming quarks ($j,k=1;\;\ell,m=1$) and even larger
otherwise. For many values of $i,j,k$, large RPV couplings are
strongly bounded from above by existing low-energy data, depending
upon sparticle masses. We therefore concentrate here on the small RPV
coupling case, where there should be no contradiction with
experimental RPV bounds over regions of MSSM parameter space that are
viable in the R-parity conserving limit.  

We now list some possible
signatures.  With small RPV couplings, one tends to have the
usual two-sparticle production via gauge couplings, in particular
strong production of gluinos and squarks.  These go through the usual
cascade decays until the LSP is reached, which decays via the RPV
coupling into Standard Model particles. Note, however, that the LSP
need not be the neutralino. For small RPV couplings there are also
large regions of parameter space where the stau is the LSP. This is
for small $m_0$ and large $m_{1/2}$. These regions also exist in the
R-parity conserving CMSSM, but are {not considered} on cosmological grounds. For
a neutralino LSP, the experimental topologies are as those in the
R-parity conserving limit, but instead of missing transverse
momentum, additional leptons and jets are obtained from the LSP decay.
For a stau LSP there are typically additional tau leptons in the final
state. Whatever the LSP, if the RPV coupling is very small, much less
than $10^{-6}$, one gets delayed LSP decays, leading to displaced
vertices, {and for the stau a potentially charged track}.

We shall first consider how the RPV couplings change the R-parity
conserving topologies in the case with a neutralino LSP and strong
sparticle production. The minimal case is di-squark production, where
each squark decays to a jet and a neutralino, which then decays via
the RPV coupling. Producing gluinos, which then decay via squarks, can
add additional jets, and there may be other particles in the chain
leading to the additional emission of leptons, weak gauge bosons or
jets. These features suggest inclusive searches, where one allows
these particles in addition to the ones listed. Since the cascades are
usually decaying on-shell particles, one can examine distributions of
invariant masses of the final state particles in order to reconstruct
bumps over the backgrounds. There is no expectation of flavour
democracy among the couplings, so flavour subtraction may be used if
backgrounds must be beaten down.

For $\lambda_{ijk}$ couplings, a neutralino LSP decays into two charged
leptons and a neutrino. Thus the inclusive minimal signature is two jets, two
leptons of one flavour, two leptons of another flavour (which may be the same
flavour as the first pair) and a moderate amount of intrinsic missing
energy. The leptons may or may not be taus for each case. 

For $\lambda'_{ijk}$ couplings, a neutralino LSP decays to a lepton
(approximately half the time into a charged lepton and half the time into a
neutrino which yields a small amount of missing transverse momentum) plus two
jets. Thus, the inclusive signal is 4 jets and two leptons. The lepton may be
either a tau or electron or muon. 

In the case of $\lambda''_{ijk}$ couplings, a neutralino LSP decays to
three jets via a virtual squark. In the case that $i,j$ or $k$ are 3,
the jets could be $t$ or $b$ quarks.  Thus, the difficult minimal
inclusive signal in this case is (from di-squark production, where
each squark decays to a jet and a neutralino LSP) eight jets with no
intrinsic missing transverse momentum.  As mentioned above, two of the
jets may be $b$ jets, and two may be replaced by decaying tops.

Next, we consider the stau--LSP case. Again the production is
dominated by squark and/or gluino pair production. The cascade decay
will typically go via a virtual lightest neutralino, which is
usually the NLSP. This
then decays to a tau and the stau LSP.  In the case that $L_3$ is
involved with the RPV superpotential coupling directly, then the stau
LSP decays directly into two particles as follows: in the case of
$\lambda'_{3jk}$ it decays into two jets, whereas in $\lambda_{ijk}$
where $i$ or $k=3$, it decays into a lepton and a neutrino.  Thus,
these topologies are covered by just assuming a neutralino LSP and
using the signature given above in the $\lambda'_{ijk}$ or
$\lambda_{ijk}$ paragraphs. 

Finally, we discuss the case that $L_3$ does {\em not} appear in the 
RPV coupling. 
In this case,
the stau decays through a four-body decay~\cite{Allanach:2003eb} into a tau and 
whatever decay products the neutralino would decay to, as specified above. 
For $\lambda_{ijk}$, the stau will decay to a tau, two charged leptons and a
neutrino. The inclusive signal is thus four charged leptons ($e,\mu$),
two taus and some missing transverse momentum.
For $\lambda'_{ijk}$ the stau decays to a tau and a lepton (neutral or
charged) and two jets. The inclusive signal is therefore 4 jets, two
leptons and two taus.
For $\lambda''_{ijk}$ the stau will decay to a
further tau plus 3 jets. The signature is thus as in the neutralino--LSP
case (8 jets) supplemented by four taus. With good tau--id the additional taus
could help search for the most difficult scenario. On the other hand,
cases where one can obtain electrons or muons from the cascade decay chain are
likely to be more feasible in terms of discovery.

\subsection{NMSSM}
The Next-to-Minimal Supersymmetric Standard Model (NMSSM) 
\cite{nmssm}
is the
simplest supersymmetric extension of the SM with a scale invariant
superpotential: the $\mu$-term in the superpotential of the MSSM is
replaced by a coupling to a gauge singlet superfield $S$, whose vev
generates automatically an effective $\mu$-term of the order of the SUSY
breaking scale. The superpotential of the NMSSM contains the terms
\begin{equation}
W_{\rm NMSSM} = \lambda S H_u H_d + \frac{\kappa}{3} S^3\; ,
\end{equation}
and the NMSSM specific soft terms are a soft mass $m_S^2$ for the scalar
components of $S$ as well as trilinear couplings $\lambda A_\lambda S
H_u H_d +  \frac{\kappa}{3} A_\kappa S^3$. Just like the MSSM, the NMSSM solves
the gauge hierarchy problem, provides a DM candidate, and leads to gauge
coupling unification. The presence of $S$ implies an extended Higgs
sector (3 neutral CP-even and 2 neutral CP-odd states) and an extended
neutralino sector (5 neutralinos including the singlino). In cases
where the LSP is singlino-like there will be significant
modifications of all sparticle decay cascades as compared to the MSSM. 
The extended Higgs sector
can include light CP-odd or CP-even states implying dominant
Higgs-to-Higgs decays, and Higgs production in sparticle decay cascades.

Because of a larger number of both superpotential and soft SUSY
breaking parameters than in the CMSSM, there are more options for
selecting unified boundary conditions at the GUT scale. In particular,
one can make less restrictive assumptions on the Higgs sector soft
terms at the GUT scale~\cite{nmssm-nuhm}.  Indeed, it is entirely
possible that the soft terms for the singlet differ from those for the
matter residing in complete SU(5) multiplets.

In the most constrained version, dubbed cNMSSM~\cite{cnmssm}, {\em
  all} soft terms are assumed to be universal at the GUT 
scale just as in the CMSSM. 
The cNMSSM has as many free parameters as the
CMSSM: the parameters $\mu$ and $B$ are replaced by $\lambda$ and
$\kappa$. With $\kappa$ being determined by $M_Z$, one is left with
\begin{equation}
\mathrm{cNMSSM:} \qquad \ m_{1/2},\ m_0,\ A_0,\ \lambda,\ \mathrm{sign}(\mu)\; .
\end{equation}
However, phenomenological constraints imply that $m_0$ and
$\lambda$ are small \cite{cnmssm}; with $A_0$ being determined by the DM
relic density, one is left with $m_{1/2}$ as the only essential free
parameter (as a function of which also $\tan\beta$ is determined
\cite{cnmssm}).

In the CNMSSM on the other hand one allows $m_S$ to differ
from $m_0$~\cite{cnmssm-lr}, in which case the free
parameters are 
\begin{equation}
\mathrm{CNMSSM:} \qquad \ m_{1/2},\ m_0,\ A_0,\ \lambda,\ \tan\beta,\  \mathrm{sign}(\mu)\; .
\end{equation}

Even more relaxed boundary conditions in the Higgs sector have been
considered. Leaving $A_\kappa$, $A_\lambda$ and $m_S$ as free
parameters and allowing as well non-universality for the Higgs soft
masses squared lead to a very much broader range of possible
phenomenologies as compared to the cNMSSM.  For example, it is easy to
find fully consistent scenarios with a singlino-like LSP {\em and}
light neutral Higgs bosons. Once $\kappa$ is determined by $M_Z$ and
$m_S$ by $\tan\beta$, the parameter space of the semi-constrained
sNMSSM can be taken as \cite{nmssm-nuhm}
\begin{equation}
\mathrm{sNMSSM:} \qquad \lambda,\  m_{1/2},\  m_0,\  m_{H_u},\  m_{H_d},\  A_0,\ 
A_\kappa,\  A_\lambda,\  \tan\beta,\  \mathrm{sign}(\mu) \; .
\end{equation}

\begin{figure*}[htb!]
\begin{center}
\resizebox{14.0cm}{!}{\includegraphics{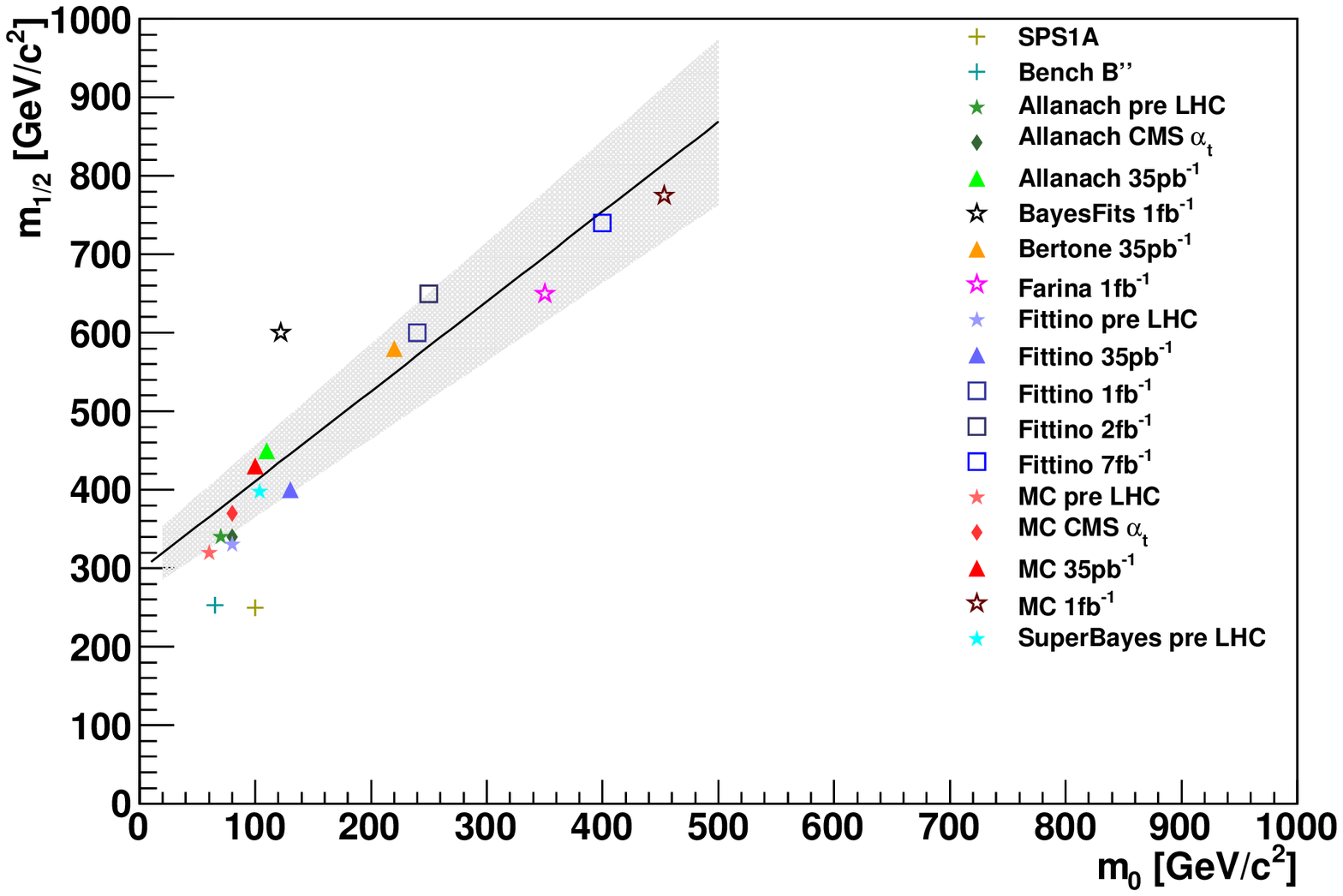}}\\
\resizebox{14.0cm}{!}{\includegraphics{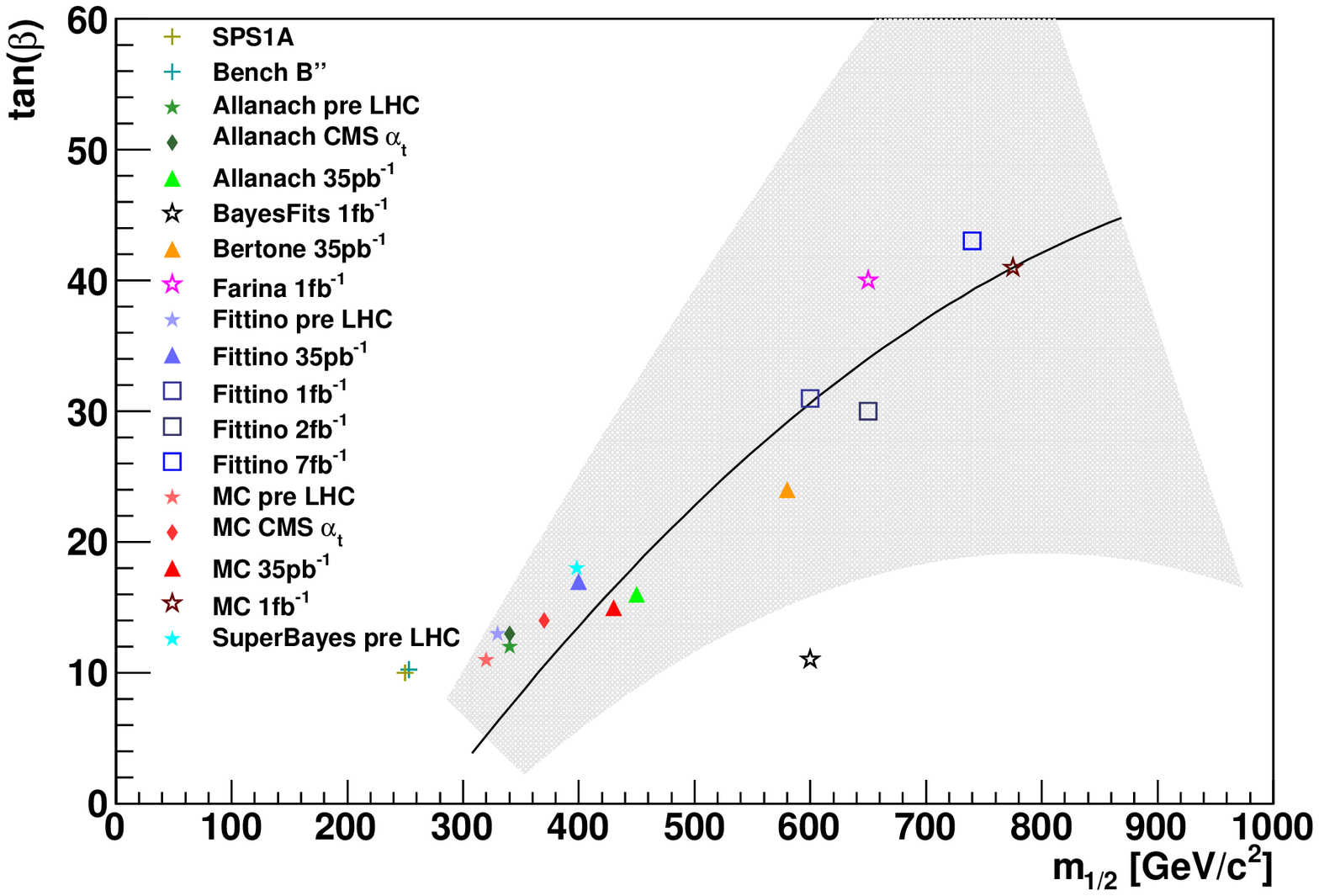}}
\end{center}
\vspace{-0.5cm}
\caption{\it CMSSM fit
  points are projected on (upper panel) the $(m_0, m_{1/2})$ plane and (lower
  panel) the $(m_{1/2}, \tb)$ plane. The best-fit points for different
  data sets are indicated by different symbols: closed stars for
  pre-LHC fits (Allanach~\cite{Allanach:2011ut},
  Fittino~\cite{Bechtle:2011dm}, MC~\cite{Buchmueller:2011aa},
  SuperBayes~\cite{Trotta:2008bp}), 
  diamonds for fits including the first SUSY searches by CMS and ATLAS
  (Allanach~\cite{Allanach:2011ut}, MC~\cite{Buchmueller:2011aa}),
  triangles including all relevant 2010 LHC data
  (Allanach~\cite{Allanach:2011wi}, Bertone~\cite{Bertone:2011nj},
  Fittino~\cite{Bechtle:2011it}, MC~\cite{Buchmueller:2011ki}),
  squares for estimates of the impacts of LHC data sets with 1, 2,
  7~\ifb\ in the absence of a SUSY
  discovery~\protect\cite{Bechtle:2011dm}, and open stars for fits
  including 1~\ifb\ of LHC data~\protect\cite{Farina:2011bh,EPS}. Also shown as
  crosses are two older benchmark points:
  SPS1a~\protect\cite{Allanach:2002nj} and a similar benchmark point
  B$^{\prime \prime}$~\protect\cite{bench}. The various symbols are also
  coded with different colours for different fitting groups, as shown
  in the legend.  The line illustrates the trend of these fits as
  stronger constraints are incorporated.  }
\label{fig:CMSSMtrajectory}
\end{figure*}


\section{Definition of Benchmark Planes, Lines and Points}
\label{sec:definitions}

One may consider two distinct approaches to the specification of
benchmarks for future SUSY searches at the LHC and elsewhere. One
possible approach is based directly on the latest available
experimental information. This has the advantage of incorporating all
the relevant experimental constraints, but also the disadvantage of
potentially becoming outmoded as the experimental constraints
evolve. The other approach is to define specific benchmark scenarios
that illustrate specific phenomenological possibilities,
as exemplified by the SPS points and lines
defined a decade ago~\cite{Allanach:2002nj}. These have the advantage that many applicable
studies have already been made, but on the other hand these points and
lines may incorporate the present experimental constraints only
partially. For example, respect of the cosmological DM density
constraint is not guaranteed, though one should bear in mind that
there are alternative sources of DM, and that in some models, e.g.,
the RPV MSSM, the DM particle might not be found among the spartners
of SM particles. 

The most commonly used benchmark scenarios for early LHC searches were defined
in~\cite{Allanach:2002nj}, where lines and points in the CMSSM, mGMSB 
and mAMSB 
were specified, as well as in the CMSSM with non-universal $m_{1/2}$.
Those Snowmass Points and Slopes (SPS) comprise six benchmark points 
(SPS~1a, SPS~1b, \ldots, SPS~5) defined within the CMSSM, one
CMSSM-like point with non-unified gaugino masses (SPS~6), two mGMSB
points (SPS~7, SPS~8) and one mAMSB point (SPS~9). The status of
those benchmark points with respect to the limits from early LHC SUSY
searches is shown in Table~\ref{tab:benchmark}. As indicated there, 
all the CMSSM points and the CMSSM-like point have now been excluded by
early LHC data. On the other hand,
the mGMSB points and the mAMSB point, which have a larger splitting
between the coloured and the colour--neutral part of the spectrum, 
continue to be valid also in view of the limits from LHC SUSY searches
with up to $\sim$2~\ifb. 

In this section we propose new benchmark planes,
lines and points, guided also by global SUSY fits. In our proposal 
we keep the SPS benchmark points of the mGMSB and mAMSB
scenarios,\footnote{We 
adopt here the definition in terms of the GUT-scale input parameters, 
while the original definition in \cite{Allanach:2002nj} was based on the 
weak-scale parameters.}
supplementing them with further points along the parameter lines. 
We propose updated parameter points and lines for the CMSSM, and we
incorporate also further SUSY scenarios that were not considered in the
SPS benchmarks. In addition, we extend our consideration of benchmark
lines and points to the NMSSM and to RPV models, which, as described above, can allow for
significantly different phenomenology as compared to the MSSM benchmarks.
Overall, by broadening the model basis, the experimental results obtained by
ATLAS and CMS can be expressed in a more general way, and their
application to a wider class of models would be feasible. Input files
for the generation of Monte Carlo events for each of the proposed
benchmark points, in the standard Les Houches accord
format~\cite{Skands:2003cj}, are
available upon request.


\begin{table}
\begin{center}
{\footnotesize
\begin{tabular}{|c|c|c|c|c|c|c|}
\hline
Benchmark point& Model scenario & \multicolumn{4}{|c|}{$\sigma/{\rm pb}$} & status\\\hline
& &A&B&C&D& ATLAS 35, 165/pb\\\hline\hline
ATLAS Limits & & 1.3 & 0.35 & 1.1 & 0.11 & \\
\hline\hline
SPS\,1a~\cite{Allanach:2002nj} & CMSSM & 2.031 & 0.933 & 1.731 & 0.418 & A,B,C,D\\\hline
SPS\,1b~\cite{Allanach:2002nj} & CMSSM & 0.120 & 0.089 & 0.098 & 0.067 & 165/pb\\\hline
SPS\,2~\cite{Allanach:2002nj}  & CMSSM & 0.674 & 0.388 & 0.584 & 0.243 & B,D\\\hline
SPS\,3~\cite{Allanach:2002nj}  & CMSSM & 0.123 & 0.093 & 0.097 & 0.067 & 165/pb\\\hline
SPS\,4~\cite{Allanach:2002nj}  & CMSSM & 0.334 & 0.199 & 0.309 & 0.144 & D      \\\hline
SPS\,5~\cite{Allanach:2002nj}  & CMSSM & 0.606 & 0.328 & 0.541 & 0.190 & D      \\\hline
SPS\,6~\cite{Allanach:2002nj}  & CMSSM~(non-universal $m_{1/2}$) & 0.721 & 0.416 & 0.584 & 0.226 & B,D    \\\hline
SPS\,7~\cite{Allanach:2002nj}  & mGMSB~($\tilde{\tau}_1$ NLSP) & 0.022 & 0.016 & 0.023 & 0.015 & allowed\\\hline
SPS\,8~\cite{Allanach:2002nj}  & mGMSB~($\tilde{\chi}^0_1$ NLSP)  & 0.021 & 0.011 & 0.022 & 0.009 & allowed\\\hline
SPS\,9~\cite{Allanach:2002nj}  & mAMSB & $0.019^*$ & $0.004^*$ &
$0.006^*$ & $0.002^*$ & allowed\\\hline
\end{tabular}}
\caption{ List of the SPS benchmark points, with their status with
  respect to the current limits from LHC SUSY searches, adapted
  from~\protect\cite{Dolan:2011ie}. For each point the columns
  labelled A,B,C and D give the cross section for each of the signal
  regions used in the 35/pb ATLAS 0-lepton
  analysis~\cite{daCosta:2011qk}. The last column shows which SPS
  points were excluded by the absence of a signal in one or more of
  these four signal regions, or by the results of a subsequent ATLAS
  0-lepton analysis using 165/pb of 2011
  data~\cite{ATLAS2011-165ipb}. ``Allowed'' reflects the lack of
  constraints from currently available results of searches with up to
  $\sim 1$~\ifb. In the mGMSB scenario the NLSP was
  taken to be stable on collider time scales.  The starred cross
  sections are computed at leading order, whereas all the other values
  are calculated at next-to-leading order
  (NLO)~\cite{Beenakker:1996ch, Beenakker:1997ut,Beenakker:1999xh}.}
\label{tab:benchmark}
\end{center}
\end{table}

\subsection{Current SUSY Fits}
\label{sec:current}

Several groups have published SUSY fits to the available data, both
before and after incorporation of the published 2010 LHC
data~\cite{Dolan:2011ie,Buchmueller:2011aa,Trotta:2008bp,Allanach:2011wi,Buchmueller:2011ki,Bechtle:2011dm,Allanach:2011ut,Bechtle:2011it,otheranalyses,Bertone:2011nj},
some fits have been made~\cite{Farina:2011bh,EPS} incorporating the preliminary CMS and ATLAS results based on $\sim 1$~\ifb\  of data,
and attempts have also
been made to foresee the evolution of these fits if the LHC does not
discover SUSY with varying amounts of integrated
luminosity~\cite{Bechtle:2011dm}. These fits are based on a variety of
statistical approaches, both Bayesian and frequentist, and the
agreement between them gives some hope for the stability and
reliability of the results.

Fig.~\ref{fig:CMSSMtrajectory} displays the results of some recent CMSSM
SUSY fits including $(g - 2)_\mu$,
in the $(m_0, m_{1/2})$ plane (above) and the $(\tb, m_{1/2})$ plane
(below). The best-fit points for different data sets are indicated by
different symbols: closed stars for
  pre-LHC fits, diamonds for fits including the first SUSY searches by
  CMS and ATLAS, triangles including all relevant 2010 LHC data,
  squares for estimates of the impacts of LHC data sets with 1, 2,
  7~\ifb\ in the absence of a SUSY
  discovery, and open stars for fits
  including 1~\ifb\ of LHC data. 
Also shown as crosses
are two older benchmark points: SPS1a~\cite{Allanach:2002nj} and a similar
benchmark point B$^{\prime \prime}$~\cite{bench}.
The various symbols are also coded with different colours for
different fitting groups. 


It is also reassuring that the pre-LHC stars lie close together, as do the
post-LHC fits. As expected, we see that the best-fit
value of $m_{1/2}$ increases monotonically with the continuing experimental
absence of SUSY. We note that the best-fit values of $m_0$ and $\tb$ also
tend to increase%
\footnote{We note in passing that, even before the LHC data, the value
  $\tb = 3$ 
often used as a default in experimental analyses was already disfavoured by all
global fits.}, which is due to the interplay of the LHC
constraints with $(g - 2)_\mu$, in particular. Reconciling the larger $m_{1/2}$
enforced by the LHC limits with $(g - 2)_\mu$ tends to require larger $\tb$, and
avoiding a charged LSP then tends to require larger $m_0$, as seen in Fig.~\ref{fig:planes}. On the other hand,
$A_0$~%
\footnote{The definition of $A_0$, as explained previously in the text,
  has to be kept in mind! Many public codes use the opposite sign
  convention for $A_0$.}%
  ~exhibits a weak tendency to decrease, though it is not tightly constrained by current data~%
  \footnote{For this reason, and because $A_0$ is of secondary importance for most of
  the spectrum, we do not discuss its behaviour in detail.}.

The fit probability, as evaluated from the $\chi^2/$dof, decreases
  monotonically, though it has not yet fallen sufficiently far to call the CMSSM
  into question. The decreasing probability is also due to the increasing tension between 
$(g-2)_\mu$ and other low-energy data on the one side and the
  non-observation of SUSY particles on the other side. 
The curved trend lines are the projections
in the $(m_0, m_{1/2})$ and $(m_{1/2}, \tb)$ planes of a least-squares
  fit to the best-fit points. 

We propose below updated benchmark planes, lines and points for the
CMSSM. However, the fact that this model has come under some pressure
from the limits of the LHC SUSY searches motivates putting a
stronger emphasis than in the past on alternative SUSY models, some of
which are discussed subsequently.

\begin{figure*}[htb!]
\resizebox{8cm}{!}{\includegraphics{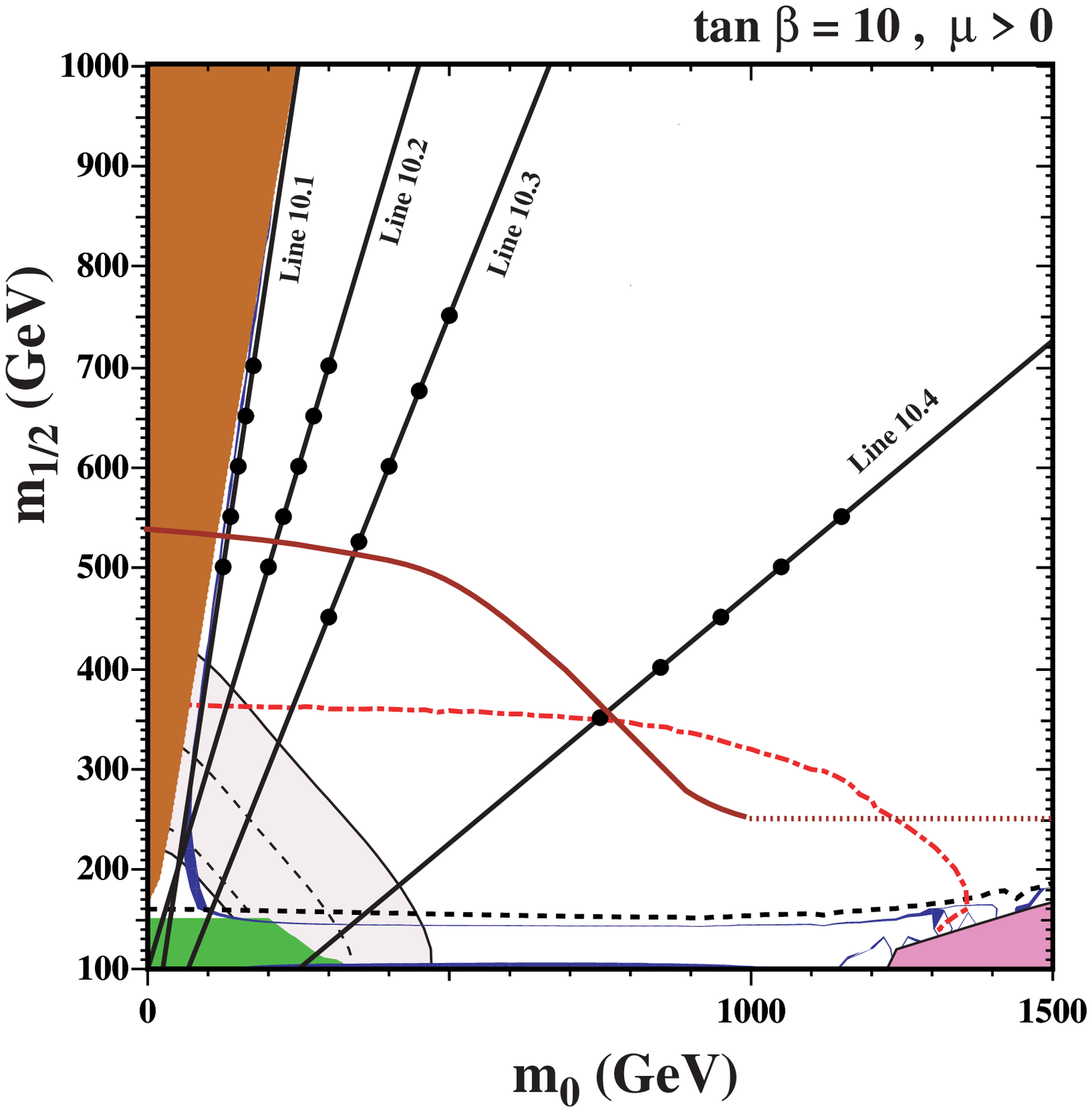}}
\resizebox{8cm}{!}{\includegraphics{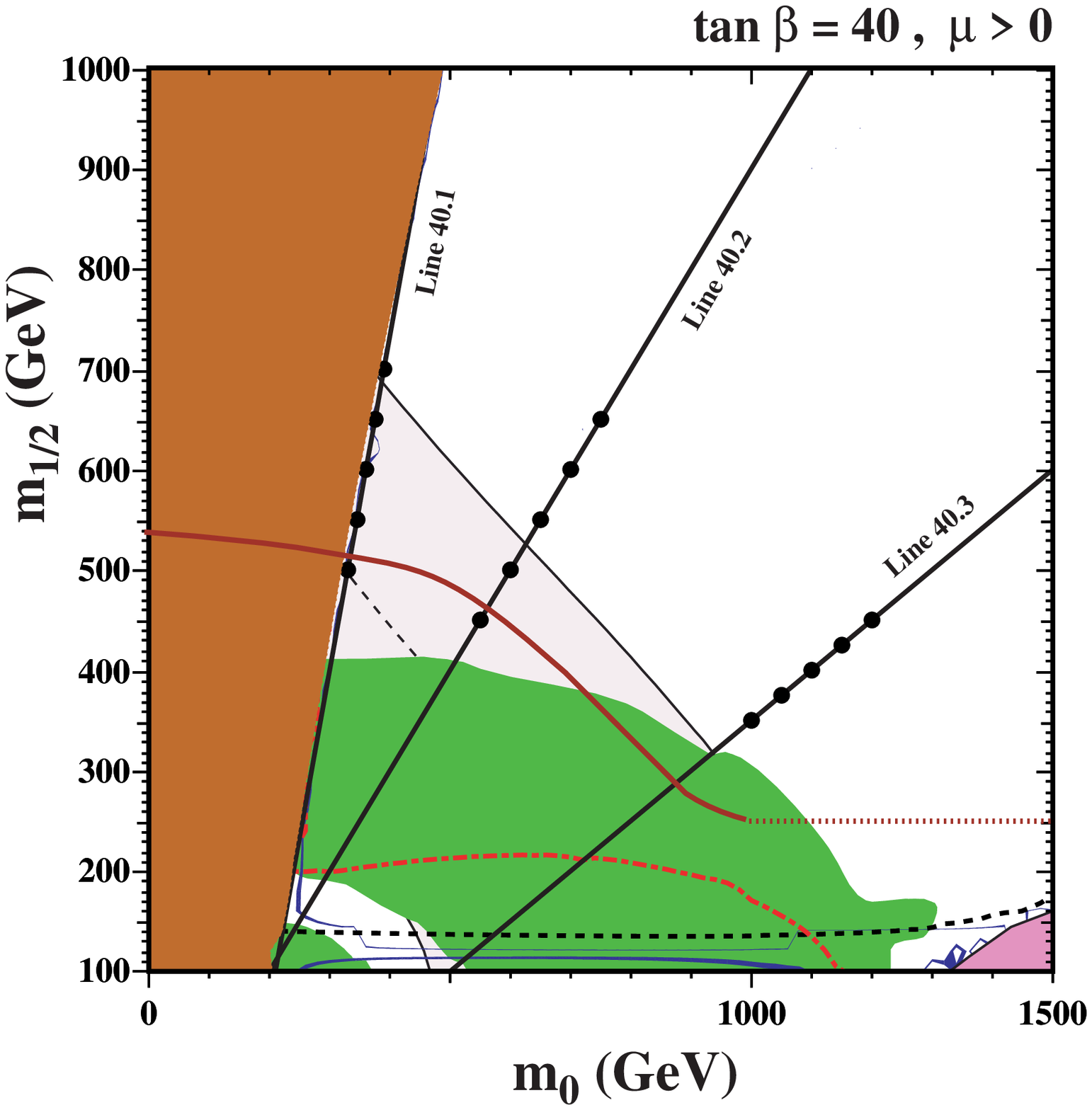}}
\vspace{-0.5cm}
\caption{\it The CMSSM $(m_0, m_{1/2})$ planes for (left) $\tan \beta = 10, \mu > 0$ and $A_0 = 0$,
and (right) $\tan \beta = 40, \mu > 0$ and $A_0 = - 500$~GeV. In the brown shaded regions at small
$m_0$ the LSP is charged, in the pink shaded regions at large $m_0$ there is no consistent
electroweak vacuum, the green shaded regions are excluded by $b \to s \gamma$, and the grey
shaded regions are favoured by $(g-2)_\mu$ at the 1- (2-)$\sigma$ level
indicated by dashed (solid)
lines. LEP searches for charginos exclude the regions below the near-horizontal black dashed lines,
LEP searches for the Higgs boson exclude the regions below the near-horizontal red dot-dashed 
lines, and LHC searches exclude the regions below the purple lines. The benchmark lines are
solid black, and the dots denote the benchmark points spaced regularly along these lines.
The dark blue strips yield the correct cold dark matter density in the CMSSM.
All experimental numbers and the corresponding references can be
found in~\cite{Buchmueller:2011ki}.}
\label{fig:planes}
\end{figure*}


\subsection{Benchmark Planes, Lines and Points in the CMSSM}
\label{sec:cmssm-planes}

Motivated by the $(g-2)_\mu$ and $b \to s \gamma$ constraints, here we fix $\mu > 0$ and propose to
consider ($m_0$, $m_{1/2}$) planes for $\tb = 10$ and $40$, which
bracket the range of $\tb$ favoured by the trend of global fits to
data before and after the start of the LHC, as shown in
Fig.~\ref{fig:planes}. 

\subsubsection{CMSSM plane I\label{sec:palin}} 
\noindent
\underline{($m_0$, $m_{1/2}$) plane:}
\begin{equation}
\mu > 0, \tan \beta=10, A_0=0
\end{equation}
Since present fits provide only weak
indications on the possible range of $A_0$, while favouring slightly
values that become increasingly negative as $m_{1/2}, m_0$ and $\tb$
increase, we propose to consider $A_0 = 0$ for the ($m_0$, $m_{1/2}$)
plane with $\tb = 10$.

\subsubsection{CMSSM plane II} 
\noindent
\underline{($m_0$, $m_{1/2}$) plane:}
\begin{equation}
\mu > 0, \tan \beta=40, A_0= - 500 \gev
\end{equation}
The value $A_0 = -500 \gev$ is proposed for the ($m_0$, $m_{1/2}$) plane
with $\tb = 40$,
so as to mirror the trend towards negative $A_0$ seen in
global fits.

\subsubsection{CMSSM lines and points}
 
As well as these planes, as also seen in Fig.~\ref{fig:planes} we propose lines
and points lying on them, that might be useful for future dedicated
detector studies and comparisons. The co-ordinates of some of the
proposed points lie close to the trend line shown in
Fig.~\ref{fig:CMSSMtrajectory}, but the lines also include points
in different regions of the $(m_0, m_{1/2})$ planes as shown in the
Tables below. The lines are chosen to illustrate a range of phenomenological
possibilities for sparticle mass hierarchies and cascade decay
branching ratios, providing a greater variety of models and SUSY
phenomenology than is covered by the CMSSM fits alone. The points are
evenly spaced along the lines, and some of the lower-mass points are
already excluded by the early LHC direct SUSY searches, whilst others
are not. At any one time, one can  define the `active point'
along any line to be the lightest one which is {\em not} ruled out by
LHC direct SUSY searches at the 95$\%$ confidence level. In this way,
if the SUSY exclusion limits advance, the points we define should
remain useful, in particular because experimental signatures are similar
for different points along the same line.

In the case of plane~I with $\tb = 10$, we propose to consider four
embedded lines, which yield different possibilities for the most
important sparticle cascade decay branching ratios.
Some properties of
the reference points, obtained from the
{\tt Suspect}~\cite{Djouadi:2002ze}/{\tt Sdecay}~\cite{Muhlleitner:2003vg}
computer codes, are shown in
Tables~\ref{tab:sfp10.1}-\ref{tab:sfp10.4}. The parameter range for
the points on each line was chosen to yield production cross sections,
summed over the various squark and gluino channels, in the range of
few--100~fb, consistent with the reach of the ongoing LHC
run.
The main qualitative features of these lines are discussed here.

{\bf Line 10.1} is defined by {\boldmath $m_0=0.25\times m_{1/2}$}, with reference
points spaced in steps of $\Delta m_{1/2}=50$~GeV.  Along this line the
gluino is heavier than the squarks, yielding {\it a priori} relatively
higher fractions of final states with fewer hadronic jets, and the
branching ratios for $\neu{2}$ decays into $\ell \equiv e/\mu$ and
$\tau$ (s)lepton flavours are relatively high, offering interesting
prospects for dilepton signatures.

\begin{table}[htb!]
\begin{center}
\caption{\underline{\bf Line 10.1}: $\tb = 10, A_0 = 0, m_0 = 0.25
  \times m_{1/2}, \Delta m_{1/2} = 50$~GeV (masses in GeV, rounded to
  5~GeV accuracy; branching
  ratios in \%). \label{tab:l10.1}}
\begin{tabular}{|c||c|c||c|c||c|c|}
\hline
 Point & $m_{1/2}$ & $m_0$ & $\mgl$ & $\langle \msq \rangle$ & BR$(\neu{2} \to {\tilde \ell}\ell)$ & 
 BR$(\neu{2} \to {\tilde \tau}\tau)$\\
\hline\hline
10.1.1 & 500 & 125   & 1145 & 1030 & 24 & 19 \\
\hline
10.1.2 & 550 & 137.5 & 1255 & 1125 & 26 & 18 \\
\hline
10.1.3 & 600 & 150   & 1355 & 1220 & 28 & 18 \\
\hline
10.1.4 & 650 & 162.5 & 1460 & 1310 & 28 & 17 \\
\hline
10.1.5 & 700 & 175   & 1565 & 1405 & 29 & 17 \\
\hline
10.1.N & $\ldots$ & $\ldots$   & $\ldots$ & $\ldots$ & $\ldots$ & $\ldots$ \\
\hline
\end{tabular}
\label{tab:sfp10.1}
\end{center}
\end{table}

{\bf Line 10.2} is defined by {\boldmath $m_0=0.5\times m_{1/2}-50$~GeV}, with reference
points spaced in steps of $\Delta m_{1/2}=50$~GeV. Along this line, 
the gluino is again heavier than the squarks, yielding {\it a priori}
relatively higher fractions of final states with fewer hadronic jets. However,
the branching ratio for $\neu{2}$ decays into the $e/\mu$ flavours is much smaller than that for the $\tau$
(s)lepton flavour.

\begin{table}[htb!]
\begin{center}
\caption{\underline{\bf Line 10.2}: $\tb = 10, A_0 = 0, m_0 = 0.5 \times
  m_{1/2} - 50~{\rm GeV}, \Delta m_{1/2} = 50$~GeV
(masses in GeV, rounded to
  5~GeV accuracy; branching ratios in \%).}
\begin{tabular}{|c||c|c||c|c||c|c|}
\hline
Point & $m_{1/2}$ & $m_0$ & $\mgl$ & $\langle \msq \rangle$ & BR$(\neu{2} \to {\tilde \ell}\ell)$ & 
BR$(\neu{2} \to {\tilde \tau}\tau)$\\
\hline\hline
10.2.1 & 500 & 200 & 1150 & 1045 & 3 & 40 \\
\hline
10.2.2 & 550 & 225 & 1255 & 1140 & 3 & 37 \\
\hline
10.2.3 & 600 & 250 & 1360 & 1235 & 2 & 34 \\
\hline
10.2.4 & 650 & 275 & 1465 & 1330 & 3  & 32 \\
\hline
10.2.5 & 700 & 300 & 1570 & 1425 & 3  & 30 \\
\hline
10.2.N & $\ldots$ & $\ldots$ & $\ldots$ & $\ldots$ & $\ldots$ & $\ldots$ \\
\hline
\end{tabular}
\label{tab:sfp10.2}
\end{center}
\end{table}

{\bf Line 10.3} is defined by {\boldmath $m_{1/2}=1.5\times m_{0}$}, with reference
points spaced in steps of $\Delta m_{0}=50$~GeV. Along this line, 
once more the gluino is heavier than the squarks. However,
the pattern of $\neu{2}$ decays is completely different, with $\neu{2} \to \neu{1} h$
decays becoming dominant, offering an interesting possible discovery channel
for the lightest MSSM Higgs boson.

\begin{table}[htb!]
\begin{center}
\caption{\underline{\bf Line 10.3}: $\tb = 10, A_0 = 0, m_{1/2} = 1.5 \times m_0, \Delta m_0= 50$~GeV
(masses in GeV, rounded to
  5~GeV accuracy; branching ratios in \%).}
\begin{tabular}{|c||c|c||c|c||c|}
\hline
Point & $m_{1/2}$ & $m_0$ & $\mgl$ & $\langle \msq \rangle$ & BR$(\neu{2} \to \neu{1} h)$\\
\hline\hline
10.3.1 & 450 & 300 & 1050 & 975 & 92 \\
\hline
10.3.2 & 525 & 350 & 1210 & 1125 & 92 \\
\hline
10.3.3 & 600 & 400 & 1370 & 1275 & 92 \\
\hline
10.3.4 & 675 & 450 & 1525 & 1420 &  92 \\
\hline
10.3.5 & 750 & 500 & 1680 & 1565 &  92 \\
\hline
10.3.N & $\ldots$ & $\ldots$ & $\ldots$ & $\ldots$ & $\ldots$ \\
\hline
\end{tabular}
\label{tab:sfp10.3}
\end{center}
\end{table}

{\bf Line 10.4} is defined by {\boldmath $m_0=2\times m_{1/2}+50$~GeV}, with reference
points spaced in steps of $\Delta m_{1/2}=50$~GeV. Along this line,
the gluino is lighter than the spartners of the lighter quark
flavours, yielding {\it a priori} relatively higher fractions of final
states with more hadronic jets.  The lighter stop and sbottom squarks
are significantly lighter than the other squarks, with the ${\tilde
  t_1}$ even lighter than the gluino though not always light enough to
overcome the kinematic restriction on direct ${\tilde g} \to {\tilde
  t_1} t$ decays. However, the branching ratios for three-body
${\tilde g} \to \neu{1} {\bar t}t, \cha{1} {\bar t}b$ decays are
dominant along this line.

\begin{table}[htb!]
\begin{center}
\caption{\underline{\bf Line 10.4}: $\tb = 10, A_0 = 0, m_0 = 2 \times m_{1/2} + 50$~GeV, $\Delta m_{1/2}= 50$~GeV
(masses in GeV, rounded to
  5~GeV accuracy; branching ratios in \%).}
\begin{tabular}{|c||c|c||c|c||c|c||c|}
\hline
 Point & $m_{1/2}$ & $m_0$ & $\mgl$ & $\langle \msq \rangle$ & $m_{\tilde t_1}$ & $m_{\tilde b_1}$ & BR$({\tilde g} \to \neu{} {\bar t}t, \cha{} {\bar t}b)$\\
\hline\hline
10.4.1 & 350 & 750 & 870 & 1040 & 720 & 915 & $53$ \\
\hline
10.4.2 & 400 & 850 & 985 & 1175 & 815 & 1035 & $71$ \\
\hline
10.4.3 & 450 & 950 & 1095 & 1310 & 910 & 1155 & $100$ \\
\hline
10.4.4 & 500 & 1050 & 1205 & 1445 & 1010 & 1275 & $100$ \\
\hline
10.4.5 & 550 & 1150 & 1320 & 1580 & 1105 & 1395 & $100$ \\
\hline
10.4.N & $\ldots$ & $\ldots$ & $\ldots$ & $\ldots$ & $\ldots$ & $\ldots$ & $\ldots$ \\
\hline
\end{tabular}
\label{tab:sfp10.4}
\end{center}
\end{table}

For the plane~II with $\tb = 40$ and
$A_0=-500$~GeV, we propose three embedded lines. 
Some properties of the reference points are shown in
Tables~\ref{tab:sfp40.1}-\ref{tab:sfp40.3}, and their main qualitative
features are discussed here.

{\bf Line 40.1} is defined by {\boldmath $m_0=0.3\times m_{1/2}+180$~GeV}, with
reference points spaced in steps of $\Delta m_{1/2}=50$~GeV. Some properties of
the reference points are shown in Table~\ref{tab:sfp40.1}. 
Along
this line, the gluino and squarks have very similar masses, and the
gluino has large branching ratios for decays into $\neu{1} {\bar t}t$
and $\cha{1} {\bar t}b$ final states. The ${\tilde \tau_1}\tau$ final
state dominates $\neu{2}$ decays, with ${\tilde \ell}\ell$ for $\ell =
e, \mu$ being suppressed because of the larger masses of the ${\tilde
  \ell}$ for this value of $\tb$.

\begin{table}[htb!]
\begin{center}
\caption{\underline{\bf Line 40.1}: $\tb = 40, A_0 = -500$~GeV, $m_0 = 0.3
  \times m_{1/2} + 180~{\rm GeV}, \Delta m_{1/2}= 50$~GeV
(masses in GeV, rounded to
  5~GeV accuracy; branching ratios in \%).}
\begin{tabular}{|c||c|c||c|c||c|c||c||c|c|}
\hline
 Point & $m_{1/2}$ & $m_0$ & $\mgl$ & $\langle \msq \rangle$ &
 $m_{\tilde t_1}$ & $m_{\tilde b_1}$ & BR$(\neu{2} \to {\tilde \tau_1}\tau/hX)$ & BR$({\tilde g} \to {\tilde t}t)$ & BR$({\tilde g} \to {\tilde b}b)$\\
\hline\hline
40.1.1 & 500 & 330 & 1155 & 1075 & 780 & 920 & 96/3 & 31 & 40 \\
\hline
40.1.2 & 550 & 345 & 1260 & 1170 & 860 & 1005 & 96/4 & 35 & 32 \\
\hline
40.1.3 & 600 & 360 & 1360 & 1260 & 935 & 1090 & 96/4 & 39 & 33 \\
\hline
40.1.4 & 650 & 375 & 1470 & 1355 & 1015 & 1175 & 94/5 & 40 &  32 \\
\hline
40.1.5 & 700 & 390 & 1570 & 1445 & 1090 & 1260 & 95/4 & 40 & 30 \\
\hline
40.1.N & $\ldots$ & $\ldots$ & $\ldots$& $\ldots$ & $\ldots$ & $\ldots$ & $\ldots$ & $\ldots$ & $\ldots$ \\
\hline
\end{tabular}
\label{tab:sfp40.1}
\end{center}
\end{table}

{\bf Line 40.2} is defined by {\boldmath $m_0= m_{1/2}+100$~GeV}, with reference
points spaced in steps of $\Delta m_{1/2}=50$~GeV. Some properties of
the reference points are shown in Table~\ref{tab:sfp40.2}, where we
indicate separately the minimum and maximum mass values of the spartners of
the light quark flavours, which bracket the gluino mass and are very
close to it. We also show the masses of the ${\tilde t_1}$
and ${\tilde b_1}$. Since these are significantly lighter, the
two-body gluino decays to $\,{\tilde t_1}{\bar t}$ and ${\tilde
  b_1}{\bar b}$ dominate, leading to $\neu{} {\bar t}t$ and $\cha{}
tb$ final states, as seen previously along {\bf Line 10.3}.
{\bf Line 40.2} also features a very large branching ratio for
$\neu{2} \to \neu{1} h$ decay.

\begin{table}[htb!]
\begin{center}
\caption{\underline{\bf Line 40.2}: $\tb = 40, A_0 = -500$~GeV, $m_0 =
  m_{1/2} + 100$~GeV, $\Delta m_{1/2} = 50$~GeV
(masses in GeV, rounded to
  5~GeV accuracy; branching ratios in \%).}
\begin{tabular}{|c||c|c||c|c||c|c||c||c|c|}
\hline
Point &
 $m_{1/2}$ & $m_0$ & $\mgl$ & $\msq$(min/max) & $m_{\tilde t_1}$ &
$m_{\tilde b_1}$ & BR$(\neu{2} \to \neu{1}h)$ & BR$({\tilde g} \to
{\tilde t}t)$ & BR$({\tilde g} \to {\tilde b}b)$\\
\hline\hline
40.2.1 & 450 & 550 & 1065 & 1060/1095 & 750 & 900 & 87 & 52 & 48 \\
\hline
40.2.2 & 500 & 600 & 1170 & 1165/1200 & 835 & 995 & 88 & 54 & 46 \\
\hline
40.2.3 & 550 & 650 & 1280 & 1270/1305 & 915 & 1090 & 89 & 56 & 44 \\
\hline
40.2.4 & 600 & 700 & 1385 & 1370/1415 & 1000 & 1185 & 90 & 57 & 42 \\
\hline
40.2.5 & 650 & 750 & 1495 & 1470/1520 & 1085 & 1280 & 91 & 58 & 41 \\
\hline
40.2.N & $\ldots$ & $\ldots $& $\ldots$ & $\ldots$ & $\ldots$ & $\ldots$ & $\ldots$ & $\ldots$ & $\ldots$ \\
\hline
\end{tabular}
\label{tab:sfp40.2}
\end{center}
\end{table}

{\bf Line 40.3} is defined by {\boldmath $m_0=2\times m_{1/2}+300$~GeV}, with
reference points spaced in steps of $\Delta m_{1/2}=50$~GeV. Some
properties of the reference points are shown in
Table~\ref{tab:sfp40.3}.  Along this line, the branching ratios for
gluino decays into light-flavour squarks and stop/sbottom 3-body
decays are comparable, as was the case along {\bf Line 10.4}. Like {\bf Line 40.2},
{\bf Line 40.3} has a large branching ratio for
$\neu{2} \to \neu{1} h^0$ decays.

\begin{table}[htb!]
\begin{center}
\caption{\underline{\bf Line 40.3}: $\tb = 40, A_0 = -500$~GeV, $m_0 = 2
  \times m_{1/2} + 300~{\rm GeV}, \Delta m_0= 50$~GeV
(masses in GeV, rounded to
  5~GeV accuracy; branching ratios in \%).}
\begin{tabular}{|c||c|c||c|c||c|c||c||c|}
\hline
 Point & $m_{1/2}$ & $m_0$ & $\mgl$ & $\langle \msq \rangle$ & $m_{\tilde t_1}$ & $m_{\tilde b_1}$ & BR$(\neu{2} \to \neu{1} h^0)$ & BR$({\tilde g} \to \neu{} {\bar t}t, \cha{} {\bar t}b)$\\
\hline\hline
40.3.1 & 350 & 1000 & 890 & 1225 & 775 & 960 & 81 &  47 \\
\hline
40.3.2 & 375 & 1050 & 950 & 1295 & 825 & 1020 & 84 &  51 \\
\hline
40.3.3 & 400 & 1100 & 1005 & 1365 & 875 & 1075 & 85 &  56 \\
\hline
40.3.4 & 425 & 1150 & 1065 & 1435 & 925 & 1135 & 86 & 61 \\
\hline
40.3.5 & 450 & 1200 & 1125 & 1505 & 975 & 1195 & 87 & 66 \\
\hline
40.3.N & $\ldots$ & $\ldots$ & $\ldots$ & $\ldots$ & $\ldots$ & $\ldots$ & $\ldots$ &  $\ldots$ \\
\hline
\end{tabular}
\label{tab:sfp40.3}
\end{center}
\end{table}

\subsection{Benchmark Planes, Lines and Points in the NUHM1 and NUHM2}
\label{sec:nuhm1placeholder}

In defining the above benchmarks in the CMSSM, we have not taken
into account the relic dark matter density constraint. As is well known,
if one assumes standard Big-Bang cosmology, 
the accuracy with which the dark matter density is determined 
essentially reduces the dimensionality of the CMSSM parameter
space by one.\footnote{This is true if one fixes other, non-CMSSM, but
  otherwise relevant parameters, in particular the mass of the top
  quark $m_t$ or $\alpha_s$. Allowing them to vary (even within their
  respective $1\,\sigma$ uncertainties) has a noticeable effect
  on the relic abundance, especially, although not only, in the regime
  of large $m_0$.
} 
For this reason, any $(m_0, m_{1/2})$ plane such as those defined above contains
only very narrow strips compatible with this constraint, shown in dark blue in Fig.~\ref{fig:planes}, flanked by regions where the
supersymmetric relic is either overdense or underdense. As a corollary,
a generic line in a CMSSM plane will yield the appropriate relic density
only at a (discrete set of) point(s). The density constraints could be relaxed
by postulating, e.g., another contribution to the dark matter in the apparently underdense
regions of parameter space, or a deviation from standard Big-Bang 
cosmology to reduce the relic density in the apparently overdense regions, or possibly by
invoking some form of $R_p$ violation that renders the LSP unstable.

The NUHM1 and NUHM2 offer one or two additional parameters that,
in general, enter into the calculation of the relic density. Hence, they offer
in principle an alternative possibility for adjusting the relic density to respect the
cosmological constraint in extended regions of a benchmark $(m_0, m_{1/2})$
plane, or along segments of the embedded benchmark lines, by adjusting
one or both of the degrees of non-universality of the SSB Higgs mass
parameters. Varying these parameters does not impact greatly the
sensitivities of the primary LHC missing-energy searches for SUSY,
though it may affect the sensitivity to flavour observables such as
BR($b \to s \gamma$) and BR($B_s \to \mu^+ \mu^-$), and certainly impacts
the predictions for and constraints imposed by Higgs searches~\cite{Buchmueller:2011aa,Buchmueller:2011ki}.

The study of these signatures lies beyond the scope of this document. For
our purposes here, it is sufficient to note that the CMSSM models discussed
above could also be taken as benchmarks for a more detailed study of
their extensions to the NUHM1 and/or NUHM2 with $M_A$ and/or
$\mu$ adjusted to fulfill the CDM constraint. With this idea in
mind, we have explored what value of $\mu$ would bring the relic density
into the WMAP range for each of the CMSSM lines and points defined above.


In the case of {\bf Line 10.1}, this exercise is unnecessary, since
the coannihilation strip for $\tb = 10$ and $A_0 = 0$ has the slope
$m_0 = 0.25  \times m_{1/2}$. In the case of {\bf Line 10.2}, a
suitable choice is~\footnote{This linear fit, and the following ones, are
  approximate, with a small dependence of the numerical values of the
  coefficients  on the code used for the renormalization-group evolution.}:
\begin{equation}
{\bf NUHM~Line~10.2:} \; \; \mbox{\boldmath$\mu \; = \; 80~ {\rm GeV}
  + 0.42 \times m_{1/2}$} ,
\label{mu102}
\end{equation}
for {\bf Line 10.3} one may choose
\begin{equation}
{\bf NUHM~Line~10.3:} \; \; \mbox{\boldmath$\mu \; = \; 74~ {\rm GeV}
  + 0.43 \times m_{1/2}$} ,
\label{mu103}
\end{equation}
and for {\bf Line 10.4} one may choose
\begin{equation}
{\bf NUHM~Line~10.4:} \; \; \mbox{\boldmath$\mu \; = \; 68~ {\rm GeV}
  + 0.44 \times m_{1/2}$} ,
\label{mu104}
\end{equation}
though the linear approximation for
$\mu$ is less accurate in this case. 

The NUHM1 modification of Line 10.2 produces some changes in the
branching ratios listed in Table~3. The $\neu{2}$ decays are dominated
by the $\tilde\tau \tau$ final state only in the case of 10.2.1 and
10.2.2. For 10.2.3 and 10.2.4 $\neu{2}$ mostly undergoes 3-body
$\neu{1}f\bar{f}$ decays mediated by an off-shell $Z^*/\gamma$. For
10.2.5 and beyond, $\neu{2}\to \neu{1}Z$. In all of these cases,
gluinos mostly decay to $\tilde{t}t$ and $\tilde{b}b$ pairs, as for the
CMSSM 10.2.X points. 
The modification of Line 10.3 has the consequence that the decay mode
$\neu{2} \to \neu{1} h$ is no longer kinematically accessible:
$\neu{2} \to \neu{1}f\bar{f}$ via an off-shell $Z^*/\gamma$ for points
1--4, while $\neu{2}\to \neu{1}Z$ for 10.3.5 and beyond. 
In the case of Line 10.4, BR$(\tilde{g}\to \tilde{t}t)=1$.

The coannihilation strip for $\tb = 40$ and $A_0 =  - 500$~GeV may
be parametrized by $m_0 = 171 {\rm GeV} + 0.318 m_{1/2}$. This is
very close to our CMSSM {\bf Line 40.1}, whose reference points provide
acceptable fits to the relic density.
In the case of {\bf Line 40.2}, the dark matter density is approximately
satisfied for
\begin{equation}
{\bf NUHM~Line~40.2:} \; \; \mbox{\boldmath$\mu \; = \; 95~ {\rm GeV} +
  0.4 \times m_{1/2}$} ,
\label{mu402}
\end{equation}
and for {\bf Line 40.3} by
\begin{equation}
{\bf NUHM~Line~40.3:} \; \; \mbox{\boldmath$\mu \; = \; 50~ {\rm GeV}
  + 0.48 \times m_{1/2}$} ,
\label{mu403}
\end{equation}
though with nonlinear deviations
at higher $m_{1/2}$.

As in the case of the $\tb=10$ reference points, the change of $\mu$
values for Lines 40.2 and 40.3 has the consequence that the decay mode
$\neu{2} \to \neu{1} h$ is no longer kinematically accessible, and
$\neu{2} \to \neu{1}f\bar{f}$, via an off-shell $Z^*/\gamma$, is the
dominant decay mode. The
gluino decays remain dominated by stop and sbottom final states, with
two-body $\tilde{g} \to \tilde{Q}Q$ ($Q=t,b$) modes for Line 40.2, and mostly
3-body $\tilde{g} \to \tilde\chi^{0(\pm)} \tilde{Q}Q^{(\prime)}$ modes for Line 40.3.


\subsection{Planes, Lines and Points in mGMSB}
\label{sec:gmsb-planes}

We propose two ($M_{\rm mess}$, $\Lambda$) planes, distinguished by
the identity of the NLSP, giving different final state topologies. 


\subsubsection{mGMSB plane I}
\label{sec:gmsb-I}
\noindent
\underline{($\Lambda$, $M_{\rm mess}$) plane:}
\begin{align}
c_{\rm grav}=1,\ N_{\rm mess} = 3,\ \tb = 15,\ \mu > 0~.
\end{align}
This plane corresponds to a stau NLSP, thus signatures
include di-taus plus missing energy plus jets. 


\subsubsection{mGMSB plane II}
\label{sec:gmsb-II}
\noindent
\underline{($\Lambda$, $M_{\rm mess}$) plane:}
This ($M_{\rm mess}$, $\Lambda$) plane is 
defined by:
\begin{align}
c_{\rm grav}=1,\ N_{\rm mess} = 1,\ \tb = 15,\ \mu > 0~.
\end{align}
This plane corresponds to a neutralino NLSP, which
may decay into a photon and a gravitino inside the detector
 if the decay is prompt enough. Thus,
signatures include jets, missing 
energy and di-photons. 

\subsubsection{mGMSB lines and points}
Here, we employ {\tt SOFTSUSY} to produce the spectra, which are successively
heavier for each point. The decay branching ratios are calculated by 
{\tt Sdecay}~\cite{Muhlleitner:2003vg} for line mGMSB2.1
and
{\tt HERWIG++-2.5.1} \cite{Bahr:2008pv} for line mGMSB1 and mGMSB2.2. 

{\bf Line mGMSB1} is defined from mGMSB plane I by $N_{\rm mess} = 3, \tb =
15, \mu > 0~, \Lambda=M_{\rm mess}/2$ with $\Delta M_{\rm mess}=10$
TeV. Table~\ref{tab:gmsb1} shows some salient features of the spectrum and
decays. 
The point mGMSB1.2 corresponds to the SPS~7 benchmark 
point~\cite{Allanach:2002nj}.

It is clear from Table~\ref{tab:gmsb1} 
that many signal events should have di-lepton
end points and missing transverse momentum from cascade decays involving the
lightest neutralinos, as well as other events involving
taus coming from the prompt stau NLSP decays into taus and gravitinos. 
\begin{table}[htb!]
\begin{center}
{
\caption{Line \underline{\bf
    mGMSB1}: $N_{\rm mess} = 3, \tb =15$, $\mu>0$, $\Delta M_{mess}=10$ TeV
  (masses in GeV, rounded to 5 GeV 
  accuracy,
  branching ratios in $\%$). $l$ stands for charged leptons of the first two
  generations.   
\label{tab:gmsb1}}
\begin{tabular}{|c||c|c|c|c|c||c|c|}
\hline
Point & $M_{mess}$/TeV & $m_{{\tilde \tau}_1}$ &$m_{\chi_1^0}$ & $m_{\tilde g}$
& $\langle m_{\tilde q} \rangle$ & BR(${\chi_1^0\rightarrow {\tilde l}_Rl}$) &
BR(${\tilde l}_R \rightarrow l \tilde G$) \\ \hline \hline
mGMSB1.1 & 70 & 110 & 140 & 840 & 785 & 56 & 3 \\  \hline
mGMSB1.2 & 80 & 125 & 165 & 950 & 885 & 59 & 8 \\ \hline
mGMSB1.3 & 90 & 140 & 185 & 1055 & 985 & 61 & 17  \\ \hline
mGMSB1.4 & 100 & 155 & 205 & 1160 & 1080 & 62 & 32 \\ \hline
mGMSB1.5 & 110 & 170 & 230 & 1265 & 1180 & 62&  47\\ \hline
mGMSB1.6 & 120 & 185 &250 & 1370 &1285  & 63 &  61\\ \hline
mGMSB1.N & $\ldots$ & $\ldots$ & $\ldots$ & $\ldots$ & $\ldots$  & $\ldots$ &
$\ldots$\\ 
\hline
\end{tabular}
}
\end{center}
\end{table}

{\bf Line mGMSB2.1} is defined from mGMSB plane II by $N_{\rm mess} = 1, \tb =
15, \mu > 0~, \Lambda=0.9 M_{mess}$ with $\Delta M_{mess}=10$
TeV. Table~\ref{tab:gmsb2} shows some salient features of the spectrum and
decays. The squark masses are larger than the gluino mass, meaning that the
muliplicity of jets will be higher, on average. Also, the possibility of 
producing $Z$s from the prompt neutralino NLSP decays for the heavier points
allows 
an additional handle on the events: one could search for jets, $\gamma$, $Z$
and missing transverse momentum. 
\begin{table}[htb!]
\begin{center}
{
\caption{Line \underline{\bf
    mGMSB2.1}: $N_{\rm mess} = 1, \tb =15$, $\mu>0$, $\Delta M_{\rm mess}=10$ TeV
  (masses in GeV, rounded to 5 GeV 
  accuracy,
  branching ratios in $\%$). 
\label{tab:gmsb2}}
\begin{tabular}{|c||c|c|c|c||c|c|}
\hline
Point & $M_{mess}$/TeV  &$m_{\chi_1^0}$ & $m_{\tilde g}$
& $\langle m_{\tilde q} \rangle$ & BR($\chi_1^0\rightarrow {\tilde G}\gamma$) &
BR($\chi_1^0 \rightarrow Z^0 \tilde G$) \\ \hline \hline
mGMSB2.1.1 & 80 & 115 & 705 & 820 & 99 & 1 \\ \hline 
mGMSB2.1.2 & 90 & 130 & 785 & 915 & 97 & 3 \\ \hline
mGMSB2.1.3 & 100 & 145 & 865 & 1010 & 95 & 5 \\ \hline
mGMSB2.1.4 & 110 & 160 & 940 & 1100 & 93 & 7 \\ \hline
mGMSB2.1.5 & 120 & 175 & 1020 & 1190 & 90  & 10 \\ \hline
mGMSB2.1.6 & 130 & 190 & 1095 & 1280 & 89 & 11 \\ \hline
mGMSB2.1.N & $\ldots$ & $\ldots$ & $\ldots$ & $\ldots$ & $\ldots$ & $\ldots$ \\
\hline
\end{tabular}
}
\end{center}
\end{table}

{\bf Line mGMSB2.2} is defined from mGMSB plane II by $N_{\rm mess} = 1, \tb =
15, \mu > 0~$, $M_{mess}=10^9$
GeV and $\Delta \Lambda=10$ TeV. 
This line has a quasi-stable neutralino NLSP, with average decay lengths of
kilometres, 
resulting in missing transverse momentum signatures. 
Table~\ref{tab:gmsb2.2} shows some salient
features of the spectrum and decays. 
The squark masses are larger than the gluino mass, meaning that the
muliplicity of jets will be higher, on average. The lightest CP-even Higgs is
produced in sparticle decay chains in association with jets, starting from
gluinos, as shown in the last two columns of the table. 
\begin{table}[htb!]
\begin{center}
{
\caption{Line \underline{\bf
    mGMSB2.2}: $N_{\rm mess} = 1, \tb =15$, $M_{\rm mess}=10^9$ GeV, $\mu>0$,
  $\Delta \Lambda=10$ TeV 
  (masses in GeV, rounded to 5 GeV 
  accuracy,
  branching ratios in $\%$). 
\label{tab:gmsb2.2}}
\begin{tabular}{|c||c|c|c|c||c|c|}
\hline
Point & $\Lambda$/TeV  &$m_{\chi_1^0}$& 
$m_{\tilde g}$ & 
$\langle m_{\tilde q} \rangle$  & 
BR($\tilde g \rightarrow \chi_2^0 jj$) &
BR($\chi_2^0\rightarrow \chi_1^0 h$)  \\ \hline \hline
mGMSB2.2.1 & 120 & 160 & 935 & 1155 & 17 & 48 \\ \hline 
mGMSB2.2.2 & 130 & 175 & 1005 & 1245 & 17 & 52 \\ \hline
mGMSB2.2.3 & 140 & 190 & 1075 & 1330 & 16  & 53 \\ \hline
mGMSB2.2.4 & 150 & 200 & 1145 & 1420 & 16  & 55 \\ \hline
mGMSB2.2.5 & 160 & 215 & 1215 & 1510 & 15  & 57 \\ \hline
mGMSB2.2.6 & 170 & 230 & 1280 & 1595 & 15 & 58 \\ \hline
mGMSB2.2.N & $\ldots$ & $\ldots$ & $\ldots$ & $\ldots$ & $\ldots$ & $\ldots$ \\
\hline
\end{tabular}
}
\end{center}
\end{table}


\subsection{Planes, Lines and Points in mAMSB}
\label{sec:amsb-planes}

We will define one ($m_{\rm aux}$, $m_0$) plane, which contains the
SPS~9 benchmark point. Furthermore we include the model line
defined in~\cite{Allanach:2002nj}.


\subsubsection{mAMSB plane I}
\label{sec:amsb-I}

This plane is adopted from SPS~9.

\noindent
\underline{($m_{\rm aux}$, $m_0$) plane:}
\begin{align}
\tb = 10, \mu > 0~.
\end{align}

\subsubsection{mAMSB line and points}
The line is defined on mAMSB plane I above.

{\bf Line mAMSB1} has $\tan \beta=10$,
$\mu>0$, $m_0=0.0075 m_{\rm
  aux}$. The discrete points have $m_{\rm aux}$ as an integer
multiple of 10 TeV, starting from 40 TeV. Some features of
the spectrum of the first few points (obtained with {\tt SOFTSUSY})
are listed in Table~\ref{tab:mamsb}. The points 
have production cross sections for gluino and squark final states
in the $6-200$fb range. 
The point mAMSB1.3 corresponds to the SPS~9 benchmark 
point~\cite{Allanach:2002nj}.

\begin{table}[htb!]
\begin{center}
\caption{\underline{\bf mAMSB}: $\tb = 10, \mu>0, m_0 = 0.0075 m_{\rm aux}$
(masses in GeV rounded to the nearest 5 GeV, branching ratios in \%).}
\begin{tabular}{|c||c|c||c|c||c|c||c|c|}
\hline
Point &
 $m_{\rm aux}$ & $m_0$ & $\mgl$ & $\langle \msq\rangle$ & $m_{\tilde t_1}$
& $m_{\tilde b_1}$ 
& BR$({\tilde g} \to \tilde t t)$ 
& BR$({\tilde g} \to \tilde b b)$ \\
\hline\hline
mAMSB1.1 & $4\times 10^{4}$ & 300 & 890  & 880 & 630 & 765 & 69 & 29 \\
\hline
mAMSB1.2 & $5\times 10^{4}$ & 375 & 1085  & 1080 & 780 & 940 & 74 & 25 \\
\hline
mAMSB1.3 & $6\times 10^{4}$ & 450 & 1280  & 1280 & 925 & 1110 &76  & 24 \\
\hline
mAMSB1.N & $\ldots$ & $\ldots$ & $\ldots$  & $\ldots$ &$\ldots$ & $\ldots$ & $\ldots$ & $\ldots$ \\
\hline
\end{tabular}
\label{tab:mamsb}
\end{center}
\end{table}

\subsection{Planes, Lines and Points in the MM-AMSB}
\label{sec:mmamsb-planes}

\subsubsection{mAMSB plane I}
The mixed modulus-anomaly-mediated SUSY breaking scenario is characterized 
by the parameters specified in Eq.~\ref{mmamsb}. The parameter $l_a$ can take 
values $0$ or $1$ depending on whether the gauge field is localized on a $D3$ or 
$D7$ brane, whereas $n_i$ can be $0$, $1$, or $1/2$ for a matter field localized on 
a $D7$ brane, a $D3$ brane, or a brane intersection. In the following we focus on a 
scenario where $l_a =1$ for all gauge fields and where $n_i = 1/2$ for all matter 
fields. Furthermore we fix $\tan\beta$ and the sign of $\mu$ in accordance with 
the benchmark scenarios in other SUSY breaking models. We thus consider a 
benchmark plane in the gravitino mass, $m_{3/2}$, and the parameter $\alpha$, 
which determines the relative weight of anomaly and gravity mediation:

\noindent
\underline{($m_{3/2}$, $\alpha$) plane:}
$
\;\;\tb = 10, \mu > 0~, n_i = 1/2, l_a = 1.
$

\subsubsection{MM-AMSB lines and points}
\label{sec:mm-amsb-lines}
For the first line in the MM-AMSB benchmark plane we fix $\alpha$ and 
vary the gravitino mass: 

\noindent
{\bf Line MM-AMSB1}: $\alpha = 5$ and $20 \le m_{3/2} \le 35$.

\noindent
This choice of $\alpha$ results in a relative size of the weak scale
gaugino mass parameters of $M_1:M_2:M_3 \simeq 1:1.2:2.2$, quite
different from the CMSSM with gaugino mass unification at the GUT
scale and a corresponding weak scale ratio of $M_1:M_2:M_3 \simeq
1:2:6$. Thus, MM-AMSB models with $\alpha \approx 5$ typically have a more compressed mass spectrum 
than the CMSSM.  Line~1 within the MM-AMSB is characterized by a SPS1a-like
phenomenology with a sequence of two-body decays and features various
mass edges. Points near line~1 furthermore provide the correct dark
matter abundance. A set of benchmark points with squark and gluino
masses in the range between 1 and 1.5~TeV is collected in
Table~\ref{tab:mm-amsb1}. The spectrum has been calculated with
\texttt{ISASUGRA}~\cite{Paige:2003mg}. It is interesting to note that 
it is possible to distinguish the MM-AMSB models on line~1 from the 
CMSSM by using information from mass edges and cross sections~\cite{mmamsb_inprep}.
\begin{table}[htb!]
\begin{center}
\caption{Line \underline{\bf MM-AMSB1}:  $20 \le m_{3/2} \le 35$ with
  $\alpha = 5$, $\tb = 10, \mu > 0~, n_i = 1/2, l_a = 1$ (sparticle masses in
  GeV rounded to the nearest 5 GeV). }
\begin{tabular}{|c||c|c||c|c|c|}
\hline
Point &
 $\alpha$ & $m_{3/2}$~ [TeV] & $\mgl$ & $\langle \msq\rangle$ & $m_{\tilde \chi^0_1}$\\
\hline\hline
MM-AMSB1.1 & 5 & 20 & 1030  & 915 & 430 \\
\hline
MM-AMSB1.2 & 5 & 25 & 1270  & 1125 & 550 \\
\hline
MM-AMSB1.3 & 5 & 30 & 1505  & 1330 & 665 \\
\hline
MM-AMSB1.4 & 5 & 35 & 1740  & 1535 & 780 \\
\hline
MM-AMSB1.N & $\ldots$ & $\ldots$ & $\ldots$  &$\ldots$ & $\ldots$ \\
\hline
\end{tabular}
\label{tab:mm-amsb1}
\end{center}
\end{table}

\noindent
{\bf Line MM-AMSB2}\, features an increasing parameter $\alpha$, resulting in a different ratio of 
weak scale mass parameters and different phenomenology, including \textit{e.g.} three-body decays. 
Along line~2 we fix $m_{3/2}$ such that the resulting squark and gluino masses
are around 1.2~TeV, 
so that the benchmark points listed in Table~\ref{tab:mm-amsb2} can be probed by upcoming LHC data 
in the near future. Note however that points 
along line~2 do not, in general, provide the correct dark matter abundance.
\begin{table}[htb!]
\begin{center}
\caption{Line \underline{\bf MM-AMSB2}: $10 \le \alpha \le 25$ with
  $\tb = 10, \mu > 0~, n_i = 1/2, l_a = 1$  (sparticle masses in
  GeV rounded to the nearest 5 GeV). $m_{3/2}$ is fixed such
  the resulting squark and gluino mass spectrum is around 1.2~TeV.}
\begin{tabular}{|c||c|c||c|c|c|c|}
\hline
Point &
 $\alpha$ & $m_{3/2}$~ [TeV] & $M_1:M_2:M_3$ & $\mgl$ & $\langle \msq\rangle$ & $m_{\tilde \chi^0_1}$\\
\hline\hline
MM-AMSB2.1 & 10 & 10 & $1:1.5:3.3$ & 1240  & 1130 & 355 \\
\hline
MM-AMSB2.2 & 15 & 6.5 & $1:1.6:3.8$ & 1280  & 1170 & 315 \\
\hline
MM-AMSB2.3 & 20 & 4.5 & $1:1.6:4.1$ & 1220  & 1115 & 280 \\
\hline
MM-AMSB2.4 & 25 & 3.5 & $1:1.6:4.3$ & 1205  & 1105 & 260 \\
\hline
MM-AMSB2.N & $\ldots$ & $\ldots$ & $\ldots$ & $\ldots$  & $\ldots$ & $\ldots$ \\
\hline
\end{tabular}
\label{tab:mm-amsb2}
\end{center}
\end{table}

\subsection{Planes, Lines and Points in the p19MSSM}
In recent fits to a phenomenological MSSM, several points which fitted
indirect data well were seen to have the lighter parts of the spectrum rather
degenerate~\cite{AbdusSalam:2011hd}. We take inspiration from these points in
order to define two simple p19MSSM planes with different phenomenology.
We also use the p19MSSM to implement `simplified models', where one is
interested in only a few sparticles giving particular signatures: the rest are
set to be heavy and therefore irrelevant at current centre of mass energies.
All parameters are defined at the electroweak scale. 

\subsubsection{p19MSSM plane I}
\label{sec:pmssm-I}
\noindent
\underline{($M_1$, $M_3=m^{\rm 1st/2nd \,gen}_{\tilde f_{L,R^*}}$) plane:}
\begin{align}
M_2=m^{\rm 3rd\,gen}_{\tilde
 f_{L,R}}=m_{\tilde e_R,\tilde \mu_R} = 2500
\gev,\ m^2_{H_u}=m^2_{H_d}=0,\ A_{t,b,\tau} = 0,\ \tb = 10, \ \mu >
0~ \end{align}
where the * in $m_{\tilde f_{R^*}}$ implies the exception for $\tilde e_R$
and $\tilde \mu_R$ whose masses are fixed.
Plane I provides significantly 
different phenomenology to previous studied MSSM scenarios. 
For low values of $M_3=m^{\rm 1st/2nd \,gen}_{\tilde f_{L,R^*}}$
  the plane allows for quasi-degenerate squarks, gluinos and neutralinos,
which are likely to result in softer jets and thus be more difficult
to detect, for any given mass of squarks and gluinos. For high values
of the same parameter the plane captures a light mass gauginos-only
scenario that could also be difficult to rule out at a hadron
collider. In the latter scenario only neutralino and/or chargino
production is possible at the LHC.

\subsubsection{p19MSSM plane II}
\label{sec:pmssm-II}
\noindent
\underline{($M_1$, $m_{\tilde l}$) plane:} 
\begin{eqnarray}
&&m_{{\tilde e}_R}=m_{{\tilde e}_L}=m_{\tilde l},\
\tb = 10,\ \mu > 0, \nonumber\\&&M_3=M_2=m^{\rm 3rd\,gen}_{\tilde f_{L,R}} =m_{\tilde
  u_R,\tilde d_R, \tilde q_L} = \mu = M_A=2500 \gev, \nonumber \\
&&A_{t,b,\tau} = 0. 
\end{eqnarray}
We are also interested in models giving di-lepton plus missing transverse
momentum signatures, in the absence of hard jets. The simplest model for this
is to have light smuons, selectrons and lightest neutralino and everything
else heavy. The only SUSY production is therefore neutralino LSP production
(possibly resulting in monojet signatures if one includes initial state
radiation) or slepton production. One can obtain missing transverse momentum
plus zero, one or two leptons.

\subsubsection{p19MSSM lines and points}
We define one p19MSSM line for each
p19MSSM parameter plane defined above. 

{\bf Line p19MSSM1} has $M_3=m^{\rm 1st/2nd \,gen}_{\tilde f_{L,R^*}}=1.2
M_1$, where 
$M_1$ is an integer multiple of 100
GeV. Some points along this line together with their gluino-squark
masses and production cross-sections are shown in
Table~\ref{tab:pmssmI}. It is currently unknown whether the lighter points have
been ruled out by LHC SUSY searches, but a high level of compression in the
MSSM spectra is known to drastically reduce the LHC experiments' acceptances,
at least with standard cuts~\cite{LeCompte:2011cn}.
These spectra have been produced with {\tt
  SOFTSUSY3.1.7}~\cite{Allanach:2009bv}. 
\begin{table}[htb!]
\begin{center}
\caption{\underline{\bf Line p19MSSM1}: $M_3=m^{\rm 1st/2nd \,gen}_{\tilde f_{L,R^*}}=1.2 M_1, \tb = 10, \mu > 0, M_2=m^{\rm 3rd\,gen}_{\tilde f_{L,R}} =m_{\tilde e_R,\tilde \mu_R} = 2500 \gev,\ m^2_{H_u}=m^2_{H_d}=0,\ A_{t,b,\tau} = 0$, $\Delta M_1=100$ GeV (masses are in units of GeV and rounded to the nearest 5 GeV). The
 cross-section given in the final column is the NLO cross-section of
 gluino and squark production at 7 TeV proton-proton collider as
 calculated by {\tt prospino}~\cite{Beenakker:1996ch,
   Beenakker:1997ut, Beenakker:1999xh}v2.1.} 
\begin{tabular}{|c||c|c|c|c|c|c|}
\hline
 Point & $M_1$ & $M_3=m^{\rm 1st/2nd \,gen}_{\tilde f_{L,R^*}}$
 & $\mgl$ & $min(\msq)$ & $M_{\neu{1}}$ & $\sigma$ (fb)\\ 
\hline\hline
p19MSSM1.1 & 300 & 360 & 435 & 450 & 280 &30080\\
\hline
p19MSSM1.2 & 400 & 480 & 570 & 525 & 330& 6123 \\ 
\hline
p19MSSM1.3 & 500 & 600 & 700 & 650 & 340 & 1576 \\ 
\hline
p19MSSM1.4 & 600 & 720 & 830 & 780 & 342 &467\\
\hline
p19MSSM1.5 & 700 & 840 & 960 & 900 & 345 &153\\
\hline
p19MSSM1.6 & 800 & 960 & 1090 & 1030 & 345 &53\\
\hline
p19MSSM1.7 & 900 & 1080 & 1215 & 1150 &345 &19\\
\hline
p19MSSM1.8 &1000& 1200 & 1340 & 1280 & 345 & 7\\
\hline
p19MSSM1.N &$\ldots$& $\ldots$ & $\ldots$ & $\ldots$ & $\ldots$ & $\ldots$\\
\hline
\end{tabular}
\label{tab:pmssmI}
\end{center}
\end{table}

{\bf Line p19MSSM2} is in p19MSSM plane II, defined along a line such that a
neutralino is the LSP: $M_1=0.75 \times m_{\tilde l}, \Delta m_{\tilde
  l}=20$ GeV. As 
can be seen from Table~\ref{tab:pmssm2},  the right-handed sleptons
are lighter than the 
left-handed sleptons. All of the points have 100$\%$ branching ratios for a
first or second generation charged slepton to go to the same flavour lepton
and a neutralino. Sneutrinos may also be produced, resulting in missing
transverse momentum, since they decay to neutrinos and
neutralinos. 
\begin{table}[htb!]
\begin{center}
\caption{\underline{\bf p19MSSM2}: $M_1 = 0.75 \times m_{{\tilde
      l}}$,
$\tb = 10, \mu > 0, M_3=M_2=m^{\rm 3rd\,gen}_{\tilde f_{L,R}} =m_{\tilde
  u_R,\tilde d_R} = \mu = M_A=2500 \gev,\  A_{t,b,\tau} = 0,$
$\Delta m_{{\tilde l}}=20$ GeV
(masses are in units of GeV and rounded to the nearest 5 GeV). The
  cross-section given in the final column is the   cross-section of neutralino
  and slepton production as calculated by {\tt
    HERWIG++-2.5.1}~\cite{Bahr:2008pv}.}  
\begin{tabular}{|c||c|c|c|c|c|c|}
\hline
Point &
 $m_{{\tilde l}}$ & $M_1$ & $M_{\chi_1^0}$ & $m_{{\tilde e}_R}$ &
$m_{{\tilde e}_L}$ & $\sigma$ (fb)\\
\hline\hline                                            
p19MSSM2.1 & 100 & 75 & 75 & 130  & 240 & 49  \\ \hline 
p19MSSM2.2 & 120 & 90 & 90 & 150  & 250 &  32 \\ \hline 
p19MSSM2.3 & 140 & 105 & 100 & 165  & 260 &28 \\ \hline 
p19MSSM2.4 & 160 & 120 & 115 & 180  & 270 &24  \\ \hline
p19MSSM2.5 & 180 & 135 & 130 & 200  & 285 &16  \\ \hline
p19MSSM2.6 & 200 & 150 & 145 & 220  & 295 &14  \\ \hline
p19MSSM2.N & $\ldots$ & $\ldots$ & $\ldots$ & $\ldots$  & $\ldots$  & $\ldots$\\
\hline
\end{tabular}
\label{tab:pmssm2}
\end{center}
\end{table}

\subsection{Benchmark Planes, Lines and Points in the RPV--CMSSM}
\label{sec:rpv-points}
To aid comparison with the
R-parity conserving case, our RPV planes are based on CMSSM plane I from
Section~\ref{sec:palin}, 
augmented by a single non-zero weak-scale RPV coupling. 
When scanning over the parameters $m_0$ and $m_{1/2}$ we
propose to include the points at low $m_0$ and high $m_{1/2}$ where
the stau is the LSP, as well as neutralino LSP points. The stau LSP decays
have been included  
in \texttt{HERWIG}~\cite{herwig} in all cases below. For potential analyses, see
also the detailed work~\cite{Desch:2010gi} on
looking for stau--LSP scenarios at the LHC. 

\subsubsection{RPV-CMSSM plane I} 
\noindent
\underline{($m_0$, $m_{1/2}$) plane:}
\begin{equation}
\lambda_{121}(M_{Z})=0.01, \mu>0, \tan \beta=10, A_0=0
\end{equation}
This plane produces a signal of four 
charged 
leptons and also missing $p_T$ from two escaping neutrinos along with additional
jets etc from squark decays. SUSY 
discovery should be easier than the corresponding CMSSM plane due to the
presence of the leptons. 
For low $m_0$ and high $m_{1/2}$, when the stau becomes the LSP, the stau 
will decay via a 4--body decay, into three jets and a tau. This is included in
\texttt{HERWIG}. 

\subsubsection{RPV-CMSSM plane II}
\noindent
\underline{($m_0$, $m_{1/2}$) plane:}
\begin{equation}
\lambda_{112}'(M_{Z})=0.001, \mu>0, \tan \beta=10, A_0=0
\end{equation}
The neutralino can
decay to $e/\nu_e+2\,$ jets. Thus for this coupling the decay of the
two neutralinos can lead to two electrons, an electron and a neutrino
or two neutrinos in the final state. The two electrons can be opposite
sign or same sign, due to the Majorana nature of the
neutralino. Furthermore there are two jets from the neutralino
decay. The stau LSP region has a four-body decay leading to two
additional $\tau$s in the final state.

\subsubsection{RPV-CMSSM plane III}  
\noindent
\underline{($m_0$, $m_{1/2}$) plane:}
\begin{equation}
\lambda''_{123}(M_{Z})=1\times10^{-4}, \tan \beta=10, A_0=0
\end{equation}
Here, we have the minimal inclusive eight-jet signature with no missing
transverse momentum.
The
search strategies proposed so far rely on the charged leptons from the cascade
decay 
\cite{Baer:1994zw,Allanach:2001if}  and the possible jet structure 
\cite{Butterworth:2009qa}

\noindent
\subsubsection{Lines}
Each RPV-CMSSM parameter plane has one associated line in parameter space. 
For two of the parameter planes, we simply use CMSSM Line 10.1, whereas
for the other, we choose a line with a stau LSP to illustrate scenarios with
additional taus in the final-state topology. We leave the RPV
parameter $\Lambda$ unchanged. 

{\bf Line RPV1} This line is defined as
$\lambda_{121}(M_{Z})=0.01, m_0=0.25 \times m_{1/2}$, with reference points
spaced in steps of  $\Delta m_{1/2}=50$ GeV. Using the di-leptons from 
cascade decays involving $\tilde\chi_2^0$ may be complicated by the presence of
electrons and muons coming from the LSP decay. 
The inclusive signal of this line is 2 jets plus four leptons, each of which may be neutral, a muon or an electron.
The muon and electron can be of either charge. Thus possible combinations of four charged leptons are
$e^+e^+e^-e^-,\,
e^+e^-\mu^+e^-,\,
e^+e^-\mu^-e^+,\,
\mu^+\mu^+e^-e^-,\,
\mu^-\mu^-e^+e^+,\,
\mu^+\mu^-e^+e^-$, with the fourth and fifth particulalry spectacular.
The basic features of the spectra of points 
RPV1.1-RPV1.5 are given in
Table~\ref{tab:l10.1}, with neutralino decays as in Table~\ref{tab:neut}.
\begin{table}[htb!]
\begin{center}
\begin{tabular}{|c|c||c|c||c|c|}
\hline
Mode & BR & Mode & BR & Mode & BR \\ \hline
$\tilde\chi_1^0 \rightarrow e^+ e^- \nu_\mu$ & 0.50 &
$\tilde\chi_1^0 \rightarrow e^+ \mu^- \nu_e$ & 0.25 &
$\tilde\chi_1^0 \rightarrow e^- \mu^+ \nu_e$ & 0.25 \\ \hline
\end{tabular}
\caption{Branching ratios of neutralino LSPs along line \underline{\bf RPV1}. They are
  independent of the mass point. Particles marked $\nu$ may also be
  anti-neutrinos. 
\label{tab:neut}}
\end{center}
\end{table}

{\bf Line RPV2}
$\lambda_{112}'(M_{Z})=10^{-3}, m_0=50$ GeV, with reference points
spaced in steps of $\Delta m_{1/2}=50$ GeV. This line has a neutralino LSP for
$m_{1/2}<350$ GeV, but a stau LSP in the selected range of reference
points $m_{1/2}\ge 400$ GeV. For the chosen value of
$\lambda_{112}'(M_{Z})=10^{-3}$, the stau decay length is in the range of
1~cm, as shown in Table~\ref{tab:stau}.
The final state signature, when the staus decay inside the
detector, is six jets plus two first-generation leptons
plus two taus. 
The charged leptons can be like-sign or opposite sign.
\begin{table}[htb!]
\begin{center}
{
\caption{Line \underline{\bf
    RPV2}: $\lambda_{112}'(M_{Z})=10^{-3}, m_0=50$ GeV, $\tan \beta=10$,
  $A_0=0$, $\mu>0$, $\Delta m_{1/2}=50$ GeV (masses in GeV, rounded to 5 GeV
  accuracy,
  branching ratios in $\%$).  Particles marked $\nu$ may also be 
  anti-neutrinos, and in the final column, the $e$ denotes electrons and
  positrons. The charge of the $\tau$ is identical to that of the $\tilde
  \tau$. 
\label{tab:stau}}
\begin{tabular}{|c||c|c|c|c|c||c|c|c|}
\hline
Point & $m_{1/2}$ & $m_{\tilde\chi_1^0}$ & $m_{{\tilde \tau}_1}$ & $m_{\tilde g}$
& $\langle m_{\tilde q} \rangle$ & $c\tau$(mm) & BR(${\tilde \tau \rightarrow \tau j j
  \nu}$) & BR(${\tilde \tau \rightarrow e j j
  \tau}$) \\ \hline \hline
RPV2.1 & 400 & 160 & 155 & 935 & 840 & 8 $\times (10^{-3}/\lambda_{112}')^2$ & 60 & 40 \\ \hline
RPV2.2 & 450 & 185 & 175 & 1040 & 935 & 10$\times (10^{-3}/\lambda_{112}')^2$ & 57 & 41 \\ \hline 
RPV2.3 & 500 & 205 & 190 & 1150 & 1030 & 11$\times (10^{-3}/\lambda_{112}')^2$ & 57 & 43 \\ \hline
RPV2.4 & 550 & 225 & 210 & 1250 & 1120 & 11$\times (10^{-3}/\lambda_{112}')^2$ & 56 & 44 \\ \hline
RPV2.N & $\ldots$ & $\ldots$ & $\ldots$ & $\ldots$ & $\ldots$ & $\ldots$ & $\ldots$ & $\ldots$ \\
\hline
\end{tabular}
}
\end{center}
\end{table}

{\bf Line RPV3}
$\lambda''_{123}(M_{Z})=1\times10^{-4}, m_0=0.25 \times m_{1/2}$, with
reference points spaced in steps of  
$\Delta m_{1/2}=50$ GeV. The inclusive signal of this model is 6 jets, 2
$b-$jets and no intrinsic missing transverse momentum. The additional leptons
coming from cascades involving 
$\tilde\chi_2^0$ will likely provide a useful handle when searching for this 
difficult SUSY scenario. The basic features of the spectra of points 
RPV3.1-RPV3.5 are given in
Table~\ref{tab:l10.1}, with neutralino decays as in Table~\ref{tab:neut2}.
\begin{table}[htb!]
\begin{center}
\begin{tabular}{|c|c||c|c|}
\hline
Mode & Branching ratio & Mode & Branching ratio\\ \hline
$\tilde\chi_1^0 \rightarrow jjb$ & 0.50 &
$\tilde\chi_1^0 \rightarrow jj \bar b$ & 0.50 \\ \hline
\end{tabular}
\caption{Branching ratios of neutralino LSPs along line \underline{\bf RPV3}. They are
  independent of the mass point. 
\label{tab:neut2}}
\end{center}
\end{table}

\subsection{Benchmark Lines and Points in the NMSSM}

As stated in Section 2.9, $m_0$ and $\lambda$ are required to be small
in the cNMSSM, leaving $m_{1/2}$ as the only essential free parameter.
Fixing $m_0 = 0$ and $\lambda = 10^{-3}$ defines a {\bf Line cNMSSM}
with points listed in Table \ref{table:cnmssm}. More properties of these
points (obtained from the NMHDECAY/NMSPEC/NMSDECAY computer codes
\cite{codes}) and proposed cuts can be found in \cite{cnmssm-lhc}. In the
cNMSSM the LSP is always singlino-like and the NLSP is a stau with
slightly larger mass. This configuration  
leads to sparticle decay cascades that are typically rich in $\tau$-leptons.

\begin{table}[!h]
\begin{center}
\begin{tabular}{|c||c|c|c|c|c|}
\hline
Point & $m_{1/2}$ &  $m_{\tilde{g}}$ & 
$m_{\tilde{u}_R}$& $m_{\tilde\chi_1^0}$ & $m_{\tilde{\tau}_1}$ \\
\hline
cNMSSM.1 & 520  & 1190 & 1045 & 142 & 147 \\
\hline
cNMSSM.2 & 600  & 1360 & 1190 & 166 & 171 \\
\hline
cNMSSM.3 & 800  & 1780 & 1545 & 225 & 229 \\
\hline
cNMSSM.4 & 1000  & 2190 & 1895 & 282 & 286 \\
\hline
cNMSSM.N & $\ldots$  & $\ldots$ & $\ldots$ & $\ldots$ & $\ldots$ \\
\hline

\end{tabular}
\end{center}
\caption{Line {\bf \underline{cNMSSM}}: $m_0 = 0$ and  $\lambda =
10^{-3}$. (Masses in GeV; gluino and squark masses rounded to 5~GeV
accuracy.)}
\label{table:cnmssm} 
\end{table}

The fact that there is also motivation for nonuniversal boundary
conditions (vs.\ the complete universality in the cNMSSM), and the fact
that $\lambda$ and $m_0$ must be so small in the cNMSSM in order to
have correct phenomenology suggests that it may be important to
consider the slightly less restricted scenarios of the {\bf CNMSSM}
and the {\bf sNMSSM} as described earlier.

A global study of the CNMSSM~\cite{cnmssm-lr} has shown that
(pre-LHC) data strongly ``pull'' $\lambda$ and $\kappa$ towards zero,
which is a decoupling limit. In this limit the phenomenology of the
model very much resembles that of the CMSSM. Nevertheless, both
$\lambda$ and $\kappa$ are allowed to be substantially different from
zero, with $\kappa\sim\lambda$. The lightest neutralino tends to be
bino dominated, and the lightest Higgs scalar is
SM-like, just as in the CMSSM. For this reason we believe that the
benchmark lines and planes defined above for the CMSSM are applicable
also to the CNMSSM. On the flip side of the coin, 
in the preferred region of parameter space it is likely to be
very challenging to distinguish between the two models at the
LHC. Additional work is needed to clarify this in more detail.

The {\bf sNMSSM}, on the other hand, leads to a vastly expanded range
of scenarios.  One scenario that exemplifies this is outlined below.
We give only a single point {\bf \underline{sNMSSM1.1}} specified by the
parameters given in Table~\ref{table:snmssm}; more will be added in the
future.

\def\gev{~{\rm GeV}}
\begin{table}[h]
\begin{center}
\begin{tabular}{|c|c|c|c|c|c|c|c|c|c|}
\hline
$\lambda$ & $m_{1/2}$ & $m_0$ & $m_{H_u}$ & $m_{H_d}$ & $A_0$ &
$A_\kappa$ & $A_\lambda$ & $\tan\beta$ & sign($\mu$) \cr
\hline
$0.406$ & $481$ & $1078$ & $5639$ & $868$ & $-1500$ & $-664$ & $-1870$
& $3.13$ & $-$ \cr
\hline
\end{tabular}
\end{center}
\caption{Point {\bf \underline{sNMSSM1.1}}: GUT scale parameters (masses
  in GeV) except for $\tan\beta$ and $\lambda$ which are given at scale $m_Z$.}
\label{table:snmssm}
\end{table}

Some of the interesting and, indeed, intriguing aspects of this point
are as follows. The LSP has a mass of $\sim 21\gev$ and is primarily
singlino-like with correct relic density from s-channel annihlation
through the singlet-like Higgs boson, $h_1$, where $m_{h_1}\simeq 49\gev$.
The SM-like Higgs, $h_2$, has $m_{h_2}\simeq 115\gev$.
The lightest CP-odd Higgs boson is singlet in
nature and has mass
$m_{a_1}\sim 21\gev$. The lightest stop has a mass of $\sim 504\gev$,
but all other squarks and the gluino have masses above a TeV.

Decay chains are quite exotic. For example, squarks and gluinos often
decay down to the $\tilde \chi_2^+$, the latter decaying 37\% of the
time into $\tilde \chi_1^+ h_2$ and $20\%$ of the time to $\tilde
\chi_1^+ Z$. The $h_2$ decays 68\% of the time to the invisible
singlinos and 20\% of the time into $h_1h_1$, where the $h_1$ decays
74\% of the time to $a_1a_1$, with $a_1$ decaying 92\% of the time to
$b\bar b$. As a result, 15\% of all $h_2$ decays and 2\% of all
left-handed squark decays result in a final state containing eight $b$ quarks.

This point serves to illustrate the huge range of scenarios that are
possible in the context of the {\bf sNMSSM}.
It will be quite challenging to discover such a scenario. The same
applies to many other ``exotic'' scenarios that are possible
within the {\bf sNMSSM} on the basis of limits on either Higgs or
supersymmetric particles.

\section{Summary}

We have proposed in this document various benchmark scenarios for
future SUSY searches at the LHC. These benchmark subspaces of the MSSM
could be useful for the presentation of experimental results, and the
specific benchmark points illustrate different possible experimental
signatures for testing and comparing detector performances. In making
our proposals, we have been guided by the impressive reaches already
demonstrated by the LHC experiments, which already exclude large
domains of the parameter spaces of some well-studied models. On the
other hand, even larger domains of parameter space remain to be
explored with the increases in luminosity and centre-of-mass energy
currently envisaged. We include in our selection of benchmarks
parameter planes that have already been partially explored as well as
embedded lines and discrete points, some of which have already been
explored by the LHC and other experiments but may serve as useful
points for comparison.  The points are defined on parameter lines such
that they can be extended to higher masses. Thus as one point becomes
excluded, a new point is already defined (the lightest one not ruled
out by an LHC experiment to 95$\%$ confidence level).

Our most detailed proposals have been for the CMSSM and related
models, but we have also made proposals in the contexts of alternative
scenarios for supersymmetry breaking. These proposals are less
developed and detailed than in the CMSSM, reflecting the fact that,
historically, the phenomenologies of these models have been studied
less. In view of the increasing pressure applied by the LHC to the
CMSSM, such alternative models may merit more phenomenological
attention in the future.  In order to encourage this, we also provide
explicit proposals for benchmark points and parameter planes in the
mGMSB, mAMSB, and the $R_p$-violating CMSSM.  We also include some
simplified models with specific collider topologies by exploiting a
19-dimensional parameterisation of a phenomenologically parameterised
weak-scale MSSM, the p19MSSM.

We note that many of the
supersymmetric signatures (for example jets plus missing transverse momentum) 
are insensitive to the presence of Higgs scalars, as long as their couplings
are not too large so as to change the supersymmetric spectrum. Thus, some features
of our planes, lines and points may apply also to the NMSSM, 
as well as to MSSM-based models. However, it is
likely that Higgs signals appearing in cascade decays would be
different in the NMSSM and singlino LSPs could extend the length of
the cascades. Further, if the LHC should discover a SM-like Higgs
above about 140 GeV, which is above the upper
limit of the MSSM but not the NMSSM, the focus of SUSY searches would
obviously shift away from the MSSM to non-minimal realisations of
SUSY. The same would apply should a relatively light Higgs boson be
observed decaying to a pair of still lighter scalars.
We have included proposals for benchmark points and a parameter line in
the NMSSM, in order to illustrate the potentially more ``exotic''
phenomenology in this model.


\section*{Acknowledgements}
We thank C. Lester, F. Moortgat, L. Pape, P. Pralavorio, G. Redlinger, D. Stuart and
A. Tapper for their input and feedback to the proposals presented
here. M.K. would like to thank J.\ Conley, L.\ Glaser and J.\ Tattersall
for helpful discussions on MM-ASMB scenarios.
H.K.D. would like to thank Tim Stefaniak and Klaus Desch for helpful
discussions.
The work of J.E. and K.A.O. was supported partly by the London
Centre for Terauniverse Studies (LCTS), using funding from the European
Research Council 
via the Advanced Investigator Grant 267352. 
U.E. acknowledges support from the French ANR LFV-CPV-LHC.
J.F.G. is supported by U.S. DOE grant No. DE-FG03-91ER40674.
The work of S.H. was supported 
in part by CICYT (grant FPA 2010--22163-C02-01) and by the
Spanish MICINN's Consolider-Ingenio 2010 Program under grant MultiDark
CSD2009-00064. The work of K.A.O. is supported in part by DOE grant DE-FG02-94ER-40823 at the University of Minnesota.




\pagebreak

\end{document}
